\documentclass[a4paper,11pt]{article}
\usepackage{rotating}
\usepackage{jheppub} 
\usepackage[utf8]{inputenc}
\usepackage{placeins} 
\usepackage{chngpage}
\usepackage{amsmath,amsfonts}
\usepackage{graphicx}
\usepackage{color}
\usepackage{xcolor}
\usepackage{relsize}
\usepackage{epstopdf}
\usepackage{hyperref}
\usepackage{mathrsfs}
\usepackage{ragged2e}
\usepackage{amssymb}
\usepackage{placeins}
\usepackage[normalem]{ ulem }
\usepackage{amsthm}
\usepackage{comment}
\usepackage{graphics}
\usepackage{caption}
\usepackage{subcaption}
\usepackage{verbatim}
\usepackage{arydshln}
\usepackage{cprotect}
\usepackage{adjustbox}

\usepackage{tabularx,booktabs}
\usepackage{multicol}
\usepackage{array,multirow}
\usepackage{booktabs}
\usepackage{colortbl}
\usepackage{float}

\def\gsim{\raise0.3ex\hbox{$\;>$\kern-0.75em\raise-1.1ex\hbox{$\sim\;$}}}
\def\lsim{\raise0.3ex\hbox{$\;<$\kern-0.75em\raise-1.1ex\hbox{$\sim\;$}}}

\makeatletter
\gdef\@fpheader{}
\vspace*{0cm}
\makeatother

\begin{document}

\title{Fermionic UV models for neutral triple gauge boson vertices}

\author[a]{Ricardo Cepedello,}
\author[b]{Fabian Esser,}
\author[b]{Martin Hirsch}
\author[b,c]{and Veronica Sanz}
\affiliation[a]{Departamento de Física Teórica y del Cosmos, 
Universidad de Granada, Campus de Fuentenueva, E-18071 Granada, Spain}
\affiliation[b]{Instituto de F\'isica Corpuscular (IFIC), 
Universidad de Valencia-CSIC, E-46980 Valencia, Spain}
\affiliation[c]{Department of Physics and Astronomy, 
University of Sussex, Brighton BN1 9QH, UK}

\emailAdd{ricepe@ugr.es}
\emailAdd{esser@ific.uv.es}
\emailAdd{mahirsch@ific.uv.es}
\emailAdd{veronica.sanz@uv.es}

\date{\today}

\abstract{ 

Searches for anomalous neutral triple gauge boson couplings (NTGCs)
provide important tests for the gauge structure of the standard
model. In SMEFT (``standard model effective field theory'') NTGCs
appear only at the level of dimension-8 operators. While the
phenomenology of these operators has been discussed extensively in the
literature, renormalizable UV models that can generate these operators
are scarce. In this work, we study a variety of extensions of the SM
with heavy fermions and calculate their matching to $d=8$ NTGC
operators. We point out that the complete matching of UV models
requires four different CP-conserving $d=8$ operators and that the
single CPC $d=8$ operator, most commonly used by the experimental
collaborations, does not describe all possible NTGC form factors.
Despite stringent experimental constraints on NTGCs, limits on the
scale of UV models are relatively weak, because their contributions
are doubly suppressed (being $d=8$ and 1-loop). We suggest a series of
benchmark UV scenarios suitable for interpreting searches for NTGCs in
the upcoming LHC runs, obtain their current limits and provide
estimates for the expected sensitivity of the high-luminosity LHC.  }

             
\maketitle


\section{Introduction\label{sec:intro}}

Searches for anomalous neutral triple gauge boson couplings (NTGCs)
provide important tests for the gauge structure of the standard model
(SM). In the SM, NTGCs vanish at tree-level, but tiny NTGCs are
generated at 1-loop order after electro-weak symmetry breaking. NTGCs
have already been searched for at the times of LEP, for a summary see
for example the review \cite{ALEPH:2013dgf}. However, limits on NTGCs
from LEP are rather weak, mainly due to the relatively moderate
$\sqrt{s}$ of only $\sqrt{s}\simeq 200$ GeV.  Considerable
improvements on these limits have been achieved already by the ATLAS
and CMS collaborations, particularly in the
$ZZ$~\cite{ATLAS:2017bcd,CMS:2020gtj} and
$Z\gamma$~\cite{ATLAS:2019gey,CMS:2015wtk} final states. It is worth
stressing that NTGCs searches at the LHC -- different from many other
BSM searches -- are not background limited at the moment. Thus, one
can expect that a sizeable increase in sensitivity will be achieved in
the high-luminosity phase of the LHC. Motivated by this expected
progress, in this paper we will study simple UV extensions of the SM,
which generate NTGCs.

In the absence of any beyond the standard model (BSM) resonances at
the LHC, the tool of choice to constrain new physics are effective
field theories (EFTs). For physics at the LHC, the phenomenologically
most important EFT is SMEFT (Standard Model EFT).  The so-called
Warsaw basis \cite{Grzadkowski:2010es} gives the complete list of
on-shell $d=6$ SMEFT operators.  For the complete Green's basis of
SMEFT at $d=6$ see, for example, \cite{Carmona:2021xtq}.  It is
straightforward to show, however, that none of the SMEFT operators at
$d=6$ will generate NTGCs.  Thus, NTGCs are one of the very few
``observables'' for which considering $d=8$ operators is essential for
the LHC.\footnote{The quotation marks on observables are deliberate.
  Experimentalists observe the scattering of two fermions to two
  (on-shell) gauge bosons and not the NTGC directly.}

A complete on-shell basis of $d=8$ SMEFT operators has been published
in \cite{Murphy:2020rsh}. None of
the $d=8$ SMEFT operators listed in \cite{Murphy:2020rsh} contains any
NTGC explicitly. Instead, the effects of NTGC operators are ``transferred'' via 
equations of motion of the field strength tensors to contact
interactions in the class $X^2D\Psi^2$, where $X$, $D$ and $\Psi$ stand 
symbolically for a field strength tensor, a covariant derivative and 
a standard model fermion. 

Operators that generate NTGCs can be found in the bosonic Green's
basis for $d=8$ SMEFT published in \cite{Chala:2021cgt} (see also
\cite{Ren:2022tvi}). The formulation of EFTs, however, contains a
certain amount of arbitrariness in the choice of operator
construction. Consider, for example, the full $d=6$ SMEFT Green's basis
in \cite{Carmona:2021xtq}: There are four operators in the class
$H^4D^2$, two of which are redundant on-shell. The {\em choice} which
of the four operators are the redundant ones is {\em arbitrary}. Worse
still, especially operators with derivatives can often be written in a
variety of equivalent ways.  This type of problem in operator choice
proliferates at $d=8$ and we will come back to a more detailed
discussion of this point in section \ref{sect:operators}.

As is well-known, however, NTGCs vanish
identically if all three gauge bosons are on-shell
\cite{Gaemers:1978hg,Gounaris:1999kf,Choudhury:2000bw}.  
In \cite{Degrande:2013kka} the author discussed the structure of
possible $d=8$ bosonic operators and identified four CP-conserving
(CPC) and three CP-violating (CPV) operators, which generate
NTGCs. All of these operators are of the type $X^2D^2H^2$, where $H$
is the SM Higgs. From the four CPC operators identified, three were
declared redundant in \cite{Degrande:2013kka}, leaving finally only
one $d=8$ CPC operator for NTGCs. This operator has become, somehow,
the standard choice for discussing NTGCs in the SMEFT context in the
literature and has been used by the ATLAS and CMS collaborations to
derive limits on the new physics scale
\cite{ATLAS:2017bcd,CMS:2020gtj}.  As we discuss in section
\ref{sect:operators}, however, the operator chosen in
\cite{Degrande:2013kka} can generate only two of the six form factors 
for NTGCs with (two) on-shell bosons.  UV models can generate four of
the six CPC vertices at the level of $d=8$ operators at 1-loop and to describe
the matching to those vertices four different operators are needed, as
we will also discuss in detail in section \ref{sect:operators}.

In this context we need to mention also the recent papers by Ellis et
al. \cite{Ellis:2019zex,Ellis:2020ljj,Ellis:2022zdw,Ellis:2023ucy}.
In these works, the authors studied $d=8$ contributions to NTGCs
in SMEFT. They pointed out that there are two more off-shell CPC
(and two more CPV) operators that will contribute to NTGCs, which have
not been considered in \cite{Degrande:2013kka}. These operators,
different from the ones in \cite{Degrande:2013kka}, are pure gauge
operators, class $X^3D^2$.  Despite this fact, however, in
\cite{Ellis:2019zex} the authors state that the contributions of these
operators to observables, such as $e^+ e^-\to Z\gamma$ must vanish in
the limit $\langle H \rangle \to 0$.  In \cite{Ellis:2023ucy} it was
then shown that contributions from these operators to the final state
$Z^*\gamma$ can be enhanced by large off-shell momenta of the initial
$Z^*$ boson.  We need to stress, however, that {\it none of our UV
  models generates this kind of pure gauge operators at 1-loop level},
as we will discuss in more detail later.

While the phenomenology of NTGCs has been explored in many papers~\cite{Renard:1981es, DeRujula:1991ufe,Baur:1992cd,Gounaris:1999kf,Choudhury:2000bw,Larios:2000ni,Rahaman:2018ujg,Hernandez-Juarez:2021mhi,Senol:2018cks,Yilmaz:2019cue,Senol:2019swu,Senol:2019qyl,Senol:2022snc}, very few UV models that can contribute to NTGCs have been
discussed in the literature. There is, for example,
\cite{Gounaris:2000tb}, where supersymmetric contributions to NTGCs
have been calculated for the MSSM.  The same authors discussed in
\cite{Gounaris:1999kf} a simple fermionic toy model. The model was
written down in the broken phase, thus the connection to SMEFT
operators is somewhat hidden, although in \cite{Gounaris:2000dn} it is
briefly mentioned the NTGCs should be generated via $d=8$ operators.
All these papers concentrate on CPC NTGCs, as \cite{Gounaris:1999kf} 
states that in fermionic models no CPV NTGC is generated at 1-loop. 

In the SM CP-violating three gauge boson couplings are zero up to at
least the two-loop level. CP-violating NTGCs within the general two
Higgs doublet model (2HDM), however, arise already at the 1-loop order
and these contributions have been calculated in
\cite{Grzadkowski:2016lpv,Belusca-Maito:2017iob}. The results of
\cite{Belusca-Maito:2017iob}, in particular, show that the expected
size of the NTGCs are at least two orders of magnitude below current
experimental sensitivities, even with the most favourable model
parameter assumptions, essentially because in the 2HDM NTGCs are
effectively generated via $d=12$ SMEFT operators at 1-loop. In
\cite{Moyotl:2015bia} the contributions to CPV NTGCs in scalar models
with several charged scalar bosons have been calculated. Results are
similar to the neutral scalar contributions in the 2HDM.

In this paper, we study UV extensions of the SM with new vector-like
fermions. These exotic fermions generate NTGCs via triangle loop
diagrams in the mass eigenstate basis, similar to the toy model
studied in \cite{Gounaris:1999kf}. Different from this early paper,
however, our calculation is carried out in the SMEFT limit. Not only
is the EFT limit more appropriate in the absence of new physics at the
LHC (so far), writing down models in the unbroken phase makes the
underlying physics also more transparent and allows us to give a more
realistic estimate of the size of the NTGCs generated. \footnote{For
  the toy model \cite{Gounaris:1999kf} states that the NTGC scales as
  $m_Z^2/M_F^2$, which is obviously not the correct scaling for a
  coupling generated from $d=8$ operators. The resolution to this
  apparent paradox is that $g_a$ -- treated as a free parameter in
  \cite{Gounaris:1999kf} -- in any realistic UV model will scale
  itself as $m_Z^2/M_F^2$, for a total of $(m_Z/M_F)^4$.}

The rest of this paper is organised as follows.  In section
\ref{sect:operators} we discuss $d=8$ operators contributing to NTGCs, 
as well as the different choices of operators for this problem. We then
examine briefly the different NTGCs themselves and how their
coefficients are related to the Wilson coefficients of the $d=8$
operators. Section \ref{sect:models} introduces a variety of fermionic
extensions of the SM. Section \ref{sect:proto} introduces a prototype
model, its matching onto the $d=8$ operators and some generalities,
which are qualitatively similar in all fermionic BSM models. Section 
\ref{sect:ManVLF} then presents several models and shows how
the Wilson coefficients for $d=8$ operators change when different
electro-weak multiplets are considered. Section \ref{sect:pheno} then
discusses briefly phenomenology, lists current experimental
constraints and outlines bounds on the Wilson coefficients. We also
comment briefly on possible future improvements. The paper then closes
with a short summary.


\section{SMEFT operators and NTGC vertices\label{sect:operators}}

In this section we will discuss SMEFT operators that generate
anomalous neutral triple gauge boson couplings (NTGCs) and give the
relation of the Wilson coefficients of these operators with the
anomalous vertices searched for by the experimental collaborations.

As it is easy to check, the gauge structure and field content of the
SM does not allow to write down any $d=6$ SMEFT operator that contains
NTGCs. As mentioned above, NTGCs vanish identically if all bosons are
on-shell. In order to identify {\em all} $d=8$ operators that generate
NTGCs, one therefore needs, in principle, to specify the complete
Green's basis of the SMEFT at $d=8$. The list of bosonic $d=8$
operators in the Green's basis can be found in \cite{Chala:2021cgt}.
We mention also \cite{Murphy:2020rsh}, where the complete $d=8$
on-shell basis including fermionic operators is listed. However, in
this basis NTGCs are removed by the use of equation of
motions. Instead, the effects of NTGCs with one off-shell boson are
contained indirectly in two-fermion two-(on-shell)-gauge boson contact
operators. No physical process is eliminated by going to the on-shell
basis, of course, but for studying aNTGCs it is more convenient to
work in the Green's basis.

Since none of the fermionic UV models will generate CP-violating
operators, we focus this discussion only on CP-conserving
NTGCs. Vertices with two on-shell and one off-shell neutral gauge
boson can be parametrised as \cite{Gounaris:1999kf,Gounaris:2000tb}:
\begin{eqnarray}\label{eq.ntgcv1}
i e \Gamma^{\alpha\beta\mu}_{ZZV}(q_1,q_2,q_3) & = & e\frac{(q_ 3^2-m_V^2)}{m_Z^2}
\Big[ f_5^V \epsilon^{\mu\alpha\beta\rho}(q_1-q_2)_{\rho} \Big],
\\ \label{eq.ntgcv2}
i e \Gamma^{\alpha\beta\mu}_{Z\gamma V}(q_1,q_2,q_3) & = & 
   e\frac{(q_ 3^2-m_V^2)}{m_Z^2}
\Big[ h_3^V \epsilon^{\mu\alpha\beta\rho} q_{2,\rho}
     + \frac{h_4^V}{m_Z^2}q_3^{\alpha}\epsilon^{\mu\beta\rho\sigma} 
         q_{3,\rho}q_ {2,\sigma} \Big] .
\end{eqnarray}
Here, $V=\gamma^*,Z^*$ is the off-shell boson. For the
complete list of vertices with CPV vertices and all bosons off-shell,
see \cite{Gounaris:1999kf,Gounaris:2000tb,Gounaris:2000dn}.  The
vertices in \eqref{eq.ntgcv1} and \eqref{eq.ntgcv2} can be derived
from the effective Lagrangian \cite{Gounaris:1999kf}:
\begin{eqnarray} \label{eq:LeffCPC}
{\cal L}_{\rm NP}^{CPC}  =  \frac{e}{2 m_Z^2} & \Big[ &
             f_5^{\gamma}(\partial^{\sigma}F_{\sigma\mu})\tilde{Z}^{\mu\beta}Z_{\beta} 
            +f_5^{Z}(\partial^{\sigma}Z_{\sigma\mu})\tilde{Z}^{\mu\beta}Z_{\beta} 
\\ \nonumber
       & & -h_3^{\gamma}(\partial^{\sigma}F_{\sigma\mu})\tilde{F}^{\mu\beta}Z_{\beta} 
            -h_3^{Z}(\partial^{\sigma}Z_{\sigma\mu})\tilde{F}^{\mu\beta}Z_{\beta} 
\\ \nonumber
          & & + \frac{h_ 4^{\gamma}}{2 m_Z^2}
         [\Box (\partial^{\sigma}F^{\rho\alpha})]\tilde{F}_{\rho\alpha}Z_{\sigma}
         +  \frac{h_ 4^{Z}}{2 m_Z^2}
    [(\Box + m_Z^2)(\partial^{\sigma}Z^{\rho\alpha})]\tilde{F}_{\rho\alpha}Z_{\sigma}
    \Big] ,
\end{eqnarray}
with the usual definition for the tensors and the dual defined as
$\tilde{X}_{\mu\nu}= 1/2 \,
\epsilon_{\mu\nu\alpha\beta}X^{\alpha\beta}$. All terms are CPC even
though naively one would expect the contrary given the presence of a
dual tensor. It can be explicitly checked as follows: Acting with $C$
on $Z_{\mu}$ gives $Z_{\mu} \to - Z_{\mu}$, while acting with $P$
leads to $Z_{0} \to Z_{0}$ and $Z_{i} \to - Z_{i}$ and $\partial_0 \to
\partial_0$, $\partial_i \to - \partial_i$ (similarly for $F_{\mu}$),
while the Levi-Civita symbol changes sign under $P$, but not $C$.

Following the shape of the Lagrangian \eqref{eq:LeffCPC} and the
output of the matching code \texttt{Matchete}
\cite{Fuentes-Martin:2022jrf}, which we used intensively in our
calculations, we define the operator basis~\footnote{Note that the
  matching in \texttt{Matchete} renders many more than these four
  operators, typically more than 100 operators at 1-loop and $d=8$ in
  any model. However, we checked explicitly that none of these contain
  NTGCs. Since there are four independent form factors (at $d=8$ and
  1-loop) our models can generate, four is also the maximal number of
  independent operators necessary to describe NTGCs.}
\begin{eqnarray} 
\label{eq:ODBB}
{\cal O}_{DB\tilde{B}} &=& i \frac{c_{DB\tilde{B}}}{\Lambda^4}
     H^{\dagger} {\tilde B_{\mu\nu}} (D^{\rho} B_{\nu\rho}) D_\mu H + {\rm h.c.} \, , 
      \\
\label{eq:ODWW}
{\cal O}_{DW\!\tilde{W}} &=& i  \frac{c_{DW\tilde{W}}}{\Lambda^4}
     H^{\dagger} {\tilde W_{\mu\nu}} (D^{\rho} W_{\nu\rho}) D_\mu H + {\rm h.c.} \, , 
     \\
\label{eq:ODWB}
{\cal O}_{DW\!\tilde{B}} &=& i  \frac{c_{DW\tilde{B}}}{\Lambda^4}
     H^{\dagger} {\tilde B_{\mu\nu}} (D^{\rho} W_{\nu\rho}) D_\mu H + {\rm h.c.} \, ,
     \\
\label{eq:ODBW}
{\cal O}_{DB\tilde{W}} &=& i  \frac{c_{DB\tilde{W}}}{\Lambda^4}
     H^{\dagger} {\tilde W_{\mu\nu}} (D^{\rho} B_{\nu\rho}) D_\mu H + {\rm h.c.} \, .
\end{eqnarray}
Different choices for the definition of Wilson coefficients are used
in the litereature, both including and excluding the scale
$\Lambda$. Here and elsewhere in this paper we choose to make the
$\Lambda$ dependence of the operators explicit.  It is straightforward
to check that these operators will generate $f_5^V$ and $h_3^V$ in
\eqref{eq:LeffCPC}, with the equation of motion of the strength
tensor, but will not contribute to $h_4^V$. Indeed contributions to
$h_4^V$ can be found in dimension 8 operators of the type $X^3D^2$
\cite{Ellis:2019zex, Ellis:2020ljj, Ellis:2022zdw, Ellis:2023ucy}, but
as it was mentioned before, these operators are not generated by any
UV model considered here at least up to 1-loop order. So we shall take
$h_4^V=0$ for the rest of the paper.\footnote{One can connect the $H$
  and the $H^{\dagger}$ in the diagrams generating the operators
  \eqref{eq:ODBB}-\eqref{eq:ODBW}.  Such a 2-loop diagram has
  the correct structure to contribute to the operators discussed in
  \cite{Ellis:2019zex, Ellis:2020ljj, Ellis:2022zdw, Ellis:2023ucy},
  in principle. Phenomenologically, however, such a 2-loop diagram
  will be irrelevant.}

Given the operator basis and the Lagrangian \eqref{eq:LeffCPC}, we can
match the Wilson coefficients to the vertices $f_5^V$ and $h_3^V$ in
\eqref{eq.ntgcv1} and \eqref{eq.ntgcv2} as follows:
\begin{eqnarray}
\label{eq:f5Z}
f_5^Z & = & \frac{v^2m_Z^2}{\Lambda^4} \, \frac{1}{c_W s_W} 
  \left[ s_W^2 c_{DB\tilde{B}} + c_W^2 c_{DW\!\tilde{W}} + \frac{1}{2} c_W s_W (c_{DW\!\tilde{B}} + c_{DB\tilde{W}}) \right],
  \\
\label{eq:f5Gam}
f_5^\gamma & = & \frac{v^2m_Z^2}{\Lambda^4} \, \frac{1}{c_W s_W} 
  \left[ c_W s_W ( - c_{DB\tilde{B}} + c_{DW\!\tilde{W}} ) - \frac{1}{2} (s_W^2 c_{DW\!\tilde{B}} - c_W^2 c_{DB\tilde{W}} )  \right],
    \\
\label{eq:h3Z}
  h_3^Z & = & \frac{v^2m_Z^2}{\Lambda^4} \, \frac{1}{c_W s_W} 
  \left[ c_W s_W ( - c_{DB\tilde{B}} + c_{DW\!\tilde{W}} ) + \frac{1}{2} (c_W^2 c_{DW\!\tilde{B}} - s_W^2 c_{DB\tilde{W}} )  \right],
  \\
\label{eq:h3Gam}
  h_3^\gamma & = & \frac{v^2m_Z^2}{\Lambda^4} \, \frac{1}{c_W s_W} 
  \left[ c_W^2 c_{DB\tilde{B}} + s_W^2 c_{DW\!\tilde{W}} - \frac{1}{2} c_W s_W ( c_{DW\!\tilde{B}} + c_{DB\tilde{W}} ) \right] ,
\end{eqnarray}
where $c_W$ and $s_W$ are the cosine and sine of the weak-mixing
angle. Here it is clear why even if operators
\eqref{eq:ODBB}-\eqref{eq:ODWB} all lead to exactly the same Lorentz
structures in the triple Z-vertex, all of them should be kept
independent, as they contribute with different prefactors to the
vertices.

A comment is needed at this point about the choice of basis compared
to the literature. For NTGC analyses as done by ATLAS
\cite{ATLAS:2017bcd} and CMS \cite{CMS:2020gtj}, it has become
customary to use the operator basis from \cite{Degrande:2013kka}.  In
this paper, a unique CPC operator is chosen:
\begin{equation}\label{eq:Cel}
\mathcal{O}_{\tilde{B}W} = i H^\dagger \tilde{B}_{\mu\nu} W^{\mu\rho}
\left\lbrace D_\rho, D^\nu \right\rbrace H + {\rm h.c.}
\end{equation}
This operator can be rewritten as:
\begin{equation}\label{eq:Cel2}
\mathcal{O}_{\tilde{B}W} = i H^\dagger \tilde{B}_{\mu\nu} W^{\mu\rho}
\Big(2 D_\rho D^\nu - [ D_\rho, D^\nu ] \Big) H + {\rm h.c.} \, ,
\end{equation}
where the commutator part, proportional to $X_\rho^{\,\nu}$, does not contain any NTGC, as can be 
easily checked. The first term in \eqref{eq:Cel2}, on the 
other hand, can be related to ${\cal O}_{DW\!\tilde{B}}$ via 
integration by parts:
\begin{equation}\label{eq:IBP}
{\cal O}_{DW\!\tilde{B}}  + {\rm h.c.} \rightarrow 
- i H^\dagger\tilde{B}_{\mu\nu} W^{\mu\rho}D_\rho D^\nu H 
- i H^\dagger D_\rho\tilde{B}_{\mu\nu} W^{\mu\rho} D^\nu H + {\rm h.c.} + ... \, ,
\end{equation}
where $...$ denotes others terms which do not contribute to the NTGCs.
As shown above, see \eqref{eq:f5Z}-\eqref{eq:h3Gam}, ${\cal
  O}_{DW\!\tilde{B}}$ contributes to all four form factors, while it
can be checked that $\mathcal{O}_{\tilde{B}W}$ will only generate the
vertices $f_5^\gamma$ and $h_3^Z$, i.e.\ the vertices associated to
$ZZ\gamma^*$ and $Z\gamma Z^*$, but not those with three Z's or two
photons ($f_5^Z = h_3^\gamma = 0$). The missing two pieces are
contained in the last term in \eqref{eq:IBP}. This fact
demonstrates that the basis of operators chosen in
\cite{Degrande:2013kka} is incomplete, as there is no reason for
$f_5^Z$ and $h_3^\gamma$ to vanish at $d=8$ 1-loop.\footnote{It is worth
  mentioning that we are not the first to realise this fact. Although
  it is not explicitly explained, the authors of \cite{Ellis:2023ucy}
  added a new CPC operator in their equation (2.2b), corresponding to the
  last piece in \eqref{eq:IBP}, to complete the basis of
  \cite{Degrande:2013kka}.}

With respect to our basis $\mathcal{O}_{\tilde{B}W} \rightarrow 
{\cal O}_{DW\!\tilde{B}} - {\cal O}_{DB\tilde{W}} \, +$ (terms without
NTGCs). After EWSB the contribution of ${\cal O}_{DW\!\tilde{B}}$ and
${\cal O}_{DB\tilde{W}}$ to $ZZZ^*$ and $Z\gamma\gamma^*$ are
identical, while for $ZZ\gamma^*$ and $Z\gamma Z^*$ they differ in a
sign and the interchange of $s_W$ and $c_W$. If one takes into account
these relations in \eqref{eq:f5Z}-\eqref{eq:h3Gam}, it can be
explicitly checked why $\mathcal{O}_{\tilde{B}W}$ contributes only to
$f_5^\gamma$ and $h_3^Z$, but not to $f_5^Z$ and $h_3^\gamma$, which
vanish exactly.

Finally, our operators do appear in the complete bosonic Green's basis
defined in \cite{Chala:2021cgt}. Their ${\cal
  O}^{(12)}_{B^2\Phi^2D^2}$ is ${\cal O}_{DB\tilde{B}}$, while ${\cal
  O}^{(9)}_{W^2\Phi^2D^2}$ is ${\cal O}_{DW\!\tilde{W}}$ and ${\cal
  O}^{(14)}_{WB\Phi^2D^2}$ is ${\cal O}_{DB\tilde{W}}$.  ${\cal
  O}_{DW\!\tilde{B}}$, on the other hand, is not in the basis of
\cite{Chala:2021cgt}, but can be obtained from a combination of ${\cal
  O}^{(14)}_{WB\Phi^2D^2}$ and ${\cal O}^{(15)}_{WB\Phi^2D^2}$.


\section{UV models and their matching to NTGC operators\label{sect:models}}

\subsection{A prototype model\label{sect:proto}}

\begin{figure}[t!]
    \centering
    \includegraphics[scale=0.5]{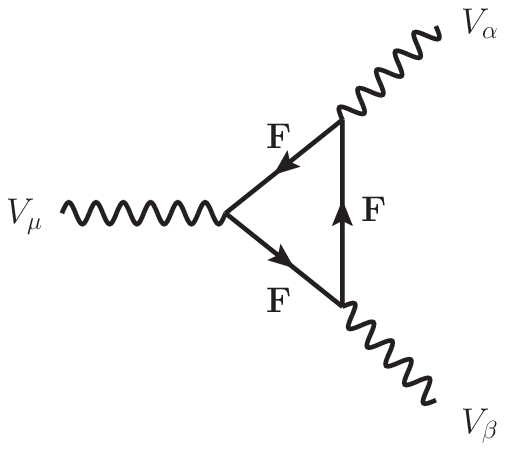}
    \includegraphics[scale=0.5]{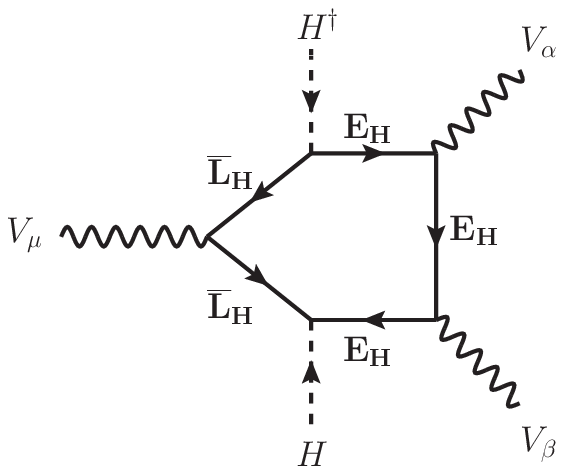}
    \caption{A fermionic triangle diagram contributing to NTGCs in the
      mass eigenstate basis, to the left. To the right, one example
      diagram for NTGCs for the prototype model in the weak basis.}
    \label{fig:ExaDiags}
\end{figure}

It is well-known \cite{Renard:1981es,Gounaris:1999kf,Gounaris:2000tb} 
that non-zero NTGCs can be generated by a fermionic triangle loop, 
see figure \ref{fig:ExaDiags} to the left. Note that this diagram 
is drawn in the mass eigenstate basis. This hides the fact that 
electro-weak symmetry must be broken, in order to generate a non-zero 
coefficient for this diagram. Small, but non-zero contributions to 
$f_5^V$ and $h_3^V$ are generated by inserting the SM fermions into 
the triangle diagram of figure \ref{fig:ExaDiags} \cite{Gounaris:2000tb}, 
but these contributions vanish rapidly with increasing $\sqrt{s}$, 
opposite to NTGCs generated by higher-dimensional operators. 

The simplest possibility to generate non-zero NTGCs therefore is to
add some heavy fermions to the SM particle content.  In principle,
only one heavy vector-like fermion is needed, but, for reasons that
will become clear soon, in our prototype model we will consider a SM
extension with one heavy lepton doublet $L_H=F_{1,2,-1/2}$ and one
heavy lepton singlet $E_H=F_{1,1,-1}$.  Here and everywhere else in
this paper, we use the notation $F$ for fermion and the subscripts
denote the representations (or charge for hypercharge) for the SM
gauge group ($SU(3)_C,SU(2)_L,U(1)_Y$).  We assume these particles to
be vector-like, i.e. their left- and right-handed Weyl spinors
transform exactly equal under the SM group.  The Lagrangian of this
model contains the terms:
\begin{eqnarray}\label{eq:LagLE}
{\cal L} & = & {\cal L}^{SM} + m_E \overline{E_H}E_H + m_L \overline{L_H}L_H 
             + ( y_{Le} \overline{L_H}e_R H + y_{lE}\overline{L}E_H H 
\\ \nonumber
            & & + y_R \overline{L_H}P_R E_H H +  y_L \overline{L_H}P_L E_H H
             + {\rm h.c.}) \, ,
\end{eqnarray}
where $L$ and $e_R$ are the SM lepton doublet and singlet, respectively. 
Note that there are two independent Yukawa couplings connecting the
heavy fields to the SM Higgs, since right- and left-handed Weyl
spinors of vector-like fermions are independent fields. In the full UV
model, $m_E$ and $m_L$ are independent parameters. However, in the EFT
limit we will consider $m_E=m_L=\Lambda$.

The $d=8$ operators we are interested in are generated via pentagon
diagrams. One example for the prototype model is shown in figure
\ref{fig:ExaDiags} to the right. This diagram is drawn in the 
gauge eigenbasis and reduces, after EWSB and rotation of the 
states to the mass eigenbasis, to the triangle diagram 
shown in the left of this figure. Note that many variations of 
this type of diagrams contribute to the total matching. 

We divide the matching of this type of loops into two sub-classes: 
Loops with two heavy particles and loops with one SM fermion and one 
heavy fermion. The former give the matching relations:
\begin{eqnarray}\label{eq:Mcbb}
  c_{DB\tilde{B}} &=&  \frac{23}{960}\frac{g_Y^2}{16 \pi^2}
      \Big(|y_L|^2 -|y_R|^2\Big),
  \\ \label{eq:Mcwb}
  c_{DB\tilde{W}} = c_{DW\tilde{B}} &= &- \frac{7}{480} 
    \frac{g_Lg_Y}{16 \pi^2}  \Big(|y_L|^2 -|y_R|^2\Big),
   \\ \label{eq:Mcww}
  c_{DW\tilde{W}} &= & \frac{1}{320} \frac{g_L^2}{16 \pi^2}
      \Big(|y_L|^2 -|y_R|^2\Big),
\end{eqnarray}
while for the heavy-light loops we find:
\begin{eqnarray}\label{eq:McbbHL}
  c_{DB\tilde{B}} = \frac{g_Y^2}{16 \pi^2}
      \Big( \frac{1}{72} |y_{Le}|^2 -\frac{7}{144}  |y_{lE}|^2\Big),
  \\ \label{eq:McwbHL}
  c_{DW\tilde{B}} = \frac{g_Lg_Y}{16 \pi^2}
      \Big( - \frac{5}{36} |y_{Le}|^2 + \frac{23}{72}  |y_{lE}|^2\Big),
   \\ \label{eq:McbwHL}
  c_{DB\tilde{W}} =  \frac{g_Lg_Y}{16 \pi^2}
      \Big( \frac{13}{36} |y_{Le}|^2 - \frac{13}{72}  |y_{lE}|^2\Big),
   \\ \label{eq:McwwHL}
  c_{DW\tilde{W}} = \frac{g_L^2}{16 \pi^2}
      \Big( \frac{1}{72} |y_{Le}|^2 + \frac{5}{144}  |y_{lE}|^2\Big).
\end{eqnarray}
Some of the coefficients for the heavy-light loops are more than a
factor of 10 larger than those for the matching with two heavy
fermions. One may therefore be tempted to think that the former will
give a more important contribution to NTGCs than the latter. However,
before drawing such a conclusion, one needs to study also constraints
on the Yukawa couplings from other operators that are generated in our
prototype model. We will come back to this point in more details in
section~\ref{sect:pheno}. For
now it is sufficient to state that constraints on $y_{Le}$ and
$y_{lE}$ from tree-level generated $d=6$ operators are in general 
so strong that contributions to NTGCs from heavy-light loops are 
negligibly small.

It is interesting to note that there is a particular limit, $y_R\equiv
y_L$, for which all heavy-heavy loops vanish exactly. The reason 
for this is that all CPC NTGCs require the presence of the totally 
anti-symmetric tensor $\epsilon^{\mu\nu\rho\sigma}$. In the calculations 
of the loops such a term can appear only from Dirac traces of the 
type: 
\begin{equation}\label{eq:TrLR}
{\rm tr}[\gamma_{\mu}\gamma_{\nu}\gamma_{\rho}\gamma_{\sigma}P_{L/R}]
 = 2 (g_{\mu\nu}g_{\rho\sigma} - g_{\mu\rho}g_{\nu\sigma} + g_{\mu\sigma}g_{\nu\rho} 
\pm  i \epsilon^{\mu\nu\rho\sigma}).
\end{equation}
If the left-handed and right-handed interactions entering a loop
calculation are exactly the same ($y_R \equiv y_L$), the terms
proportional to $P_{L/R}$, cancel each other and the coefficients for
the NTGCs vanish identically. 

We need to briefly comment also on the recent papers
\cite{Ellis:2019zex,Ellis:2020ljj,Ellis:2022zdw,Ellis:2023ucy}.  As
discussed in the introduction, in these papers it was pointed out that
there are four more $d=8$ operators that contribute to (off-shell)
NTGCs. The two CP-conserving operators in this new class can be
written as:\footnote{These operators correspond to some combination of
  operators ${\cal O}^{(3)}_{W^2BD^2}$ and ${\cal O}^{(4)}_{W^2BD^2}$
  in \cite{Chala:2021cgt}.}
\begin{eqnarray}\label{eq:Oplus}
{\cal O}_{G+} = \tilde{B}_{\mu\nu}W^{\mu\rho}
         (D_{\rho}D_{\lambda}W^{\nu\lambda} + D^{\nu}D^{\lambda}W_{\lambda\rho}),
\\ \label{eq:Ominus}
{\cal O}_{G-} = \tilde{B}_{\mu\nu}W^{\mu\rho}
         (D_{\rho}D_{\lambda}W^{\nu\lambda} - D^{\nu}D^{\lambda}W_{\lambda\rho}).
\end{eqnarray}
Such operators are never generated at 1-loop in any of our models, 
see footnote on page 5. The reason
for this is straightforward: Fermions can contribute to bosonic SMEFT
operators in the unbroken phase only if they have a vector-like mass
term. This requirement guarantees that all gauge interactions of these
exotics are left-right symmetric (before EWSB) and thus, all
contributions to the loops proportional to the anti-symmetric
tensor cancel exactly.  This is true not only for our prototype
model, but for all models built on heavy fermions.

\subsection{Many variations of heavy fermion models\label{sect:ManVLF}}

Having discussed one particular model with heavy vector-like fermions
in some detail, in this subsection we will discuss a number of
variants.  
We will concentrate on models with colour singlets only. The 
reason for this is simple. While vector-like quark loops have 
a colour factor of 3, resulting in potentially larger coefficients 
for the NTGC operators, constraints on coloured particle masses from LHC 
searches are much stronger than those for uncoloured fermions.
The simplest fermionic models that will generate NTGCs add
only one vector-like fermion to the SM. Two examples were already
briefly discussed above. Since we need to couple the heavy fermions to
the SM Higgs boson, there are in total only six leptophilic model
variants of this kind: $L_H$ and $E_H$, discussed in the previous
subsection, plus $F_{1,1,0}\equiv N_R$, $F_{1,3,0}\equiv \Sigma$,
$F_{1,2,-3/2}$ and $F_{1,3,-1}$. For completeness, we write down the
Yukawa couplings of these fields to the SM leptons:
\begin{eqnarray}\label{eq:LagSPM}
{\cal L}^{\rm Yuk} = y_{\nu}\overline{N_R}LH +  y_{\Sigma}\overline{\Sigma}LH
      + y_{F_3}\overline{F_{1,3,-1}}LH^{\dagger} 
      + y_{F_2}\overline{F_{1,2,-3/2}}e_RH^{\dagger} + h.c. 
\end{eqnarray}
The terms for $L_H$ and $E_H$ have already been given in
\eqref{eq:LagLE} and we do not write down the vector-like 
mass terms for the heavy fields for brevity. 

\begin{table}[h]
\centering
\begin{tabular}{|l|>{$}l<{$}|>{$}c<{$}|>{$}c<{$}|>{$}c<{$}|>{$}c<{$}|}
\hline
Model & \text{Particles} & \tilde{c}_{DB\tilde{B}} & \tilde{c}_{DW\tilde{B}} & \tilde{c}_{DB\tilde{W}} & \tilde{c}_{DW\tilde{W}} \\
\hline
\hline
Type-I & (L, N_R) & \frac{5}{144} & \frac{5}{72} & \frac{5}{72} & \frac{5}{144} \\
\hline
Type-III & (F_{1,3,0}, L) & -\frac{5}{48 \sqrt{3}} & \frac{7}{24 \sqrt{3}} & -\frac{17}{24 \sqrt{3}} & \frac{1}{16 \sqrt{3}}
\\
\hline
MLH & (L_H, e_R) & \frac{1}{72} & -\frac{5}{36} & \frac{13}{36} & \frac{1}{72} \\ 
\hline
MEH & (L, E_H) & -\frac{7}{144} & \frac{23}{72} & -\frac{13}{72} & \frac{5}{144} \\
\hline
MFD & (F_{1,2,-\frac{3}{2}}, e_R) & \frac{11}{72} & -\frac{7}{36} & \frac{11}{36} & -\frac{1}{72} \\
\hline
MFT & (F_{1,3,1}, L) & -\frac{7}{48 \sqrt{3}} & \frac{1}{8 \sqrt{3}} & -\frac{\sqrt{3}}{8} & -\frac{1}{16 \sqrt{3}} \\
\hline
\hline
\end{tabular}
\caption{The $\tilde{c}_{DAB}$ for models with only one heavy fermion.
  The first column gives the model name, the second the particle
  combination that enters the calculation of the loop for the NTGC
  operators. Column 3-6 give the matching coefficients.}
\label{tab:ciHL}
\end{table}

We have calculated the matching to the NTGC operators for all six 
possible models and summarise them in table \ref{tab:ciHL}. Here, only the 
coefficients of $\tilde{c}_{DAB}$ are given and we define for more 
compact notation in the tables  $\tilde{c}_{DAB}$ as:
\begin{equation}\label{eq:defct}
  c_{DAB} = \frac{1}{16\pi^2} g_A g_B |Y|^2 \tilde{c}_{DAB}, 
\end{equation}
where $Y$ denotes symbolically the corresponding Yukawa coupling of
the model (familiy indices suppressed) and $g_Ag_B$ stands for $g_Y^2$,
$g_Lg_Y$, $g_Lg_Y$ and $g_L^2$ for $AB=B\tilde{B},W\tilde{B},B\tilde{W}$, 
$W\tilde{W}$.

As the table shows again, loops contributing to the $d=8$ operators
with only one heavy field in the loop have coefficients which are
considerably larger than those for models with two heavy fields.  The
maximal variation for the vertices $f_5^V$ for the different
$\tilde{c}_{AB}$ of table \ref{tab:ciHL}, however, is less than a
factor 4. This implies that the limits on the scale $\Lambda$ that can
be derived in the different cases should vary less then roughly 
$40 \%$ and there is no model variant in this single field extension
class that will give significantly more (or less) stringent
constraints.

Two more comments are in order. Firstly, different from the case of
Yukawa couplings connecting two heavy fields, for the single field
models the Yukawa couplings are constrained from the fact that several
$d=6$ SMEFT operators are generated at tree-level.  These constraints
imply that for $\Lambda \sim {\cal O}(100 \hskip1mm {\rm GeV})$ the
Yukawas must be (much) smaller than $1$ and none of the single field
extensions can be constrained significantly from the non-observation
of NTGCs. This will be further discussed in section~\ref{sect:pheno}.

Secondly, for $N_R$ and $\Sigma$ there is another, much more powerful
bound, that we need to discuss it briefly. Fermions with quantum
numbers $F_{C,X,0}$ can be self-conjugate fields, i.e. Majorana
fermions.  If $N_R$ and $\Sigma$ are Majorana fermions, however, they
will generate neutrino masses via the type-I and type-III seesaw
mechanisms. In that case, a typical estimate for the Yukawa couplings
would be (very roughly) $Y \simeq \sqrt{(m_{\nu}\Lambda)/v^2} \simeq 3
\times 10^{-7}$ for $m_{\nu} =\sqrt{\Delta(m^2_{\rm Atm})}$ and
$\Lambda=100$ GeV. For seesaw models, the matching results we show 
in table \ref{tab:ciHL}, are therefore only of academic interest. 

There is, on the other hand, also the possibility to assume that
lepton number is conserved. One could have a vector-like field $N$,
with a conserved lepton number, if $N=(N_L,N_R)^T$ and $N_L \ne N_R^c$
-- just like for all other vector-like fermions.  In this case, the
small neutrino masses would not impose a constraint on the Yukawa
couplings $y_{\nu}$ (similar arguments hold for $y_{\Sigma}$).  Such
an assumption is not completely unreasonable. Neutrino mass models of
this kind exist in the literature, the inverse seesaw
\cite{Mohapatra:1986bd} probably being the most prominent example.
Note, however, that constraints on model parameters from $d=6$
operators still apply even in this case.

Let us now turn to the case with two (or more) exotic fermions. In
principle, fermionic $SU(2)$ multiplets of any dimension (larger than
1) can couple to $\tau^aW^{\mu,a}$, thus for the loop-generated NTGCs
an infinite tower of models can be written down in theory. Similarly,
one can write down a series of fermionic multiplets with ever larger
hypercharges, all of which couple to $B^{\mu}$.  However, one can
formulate a number of constraints that will limit the number of models
that make sense phenomenologically. For example, adding a single
fermionic ${\bf 7}$-tet to the SM will cause a Landau pole in $g_2$
below the GUT scale \cite{Cirelli:2005uq}.  Models with several
multiplets (and with higher hypercharges) are even more constrained by
this argument. In practice, we have therefore limited the models that
we study to models up to quintuplets in
$SU(2)$ and with a maximum hypercharge $Y$ of $Y=4$.

To generate NTGCs with models containing two VLFs, the quantum numbers
of the two fields have to satisfy two relations: (i) The product of
the two $SU(2)$ multiplets must contain a doublet and (ii) the
difference in hypercharge between the two VLF must be $|\Delta Y|=1/2$.
These conditions can be understood trivially, since a Yukawa coupling
of the form $Y\overline{F_1}F_2H$ is needed to generate the NTGC
operators.

We have implemented a number of such models in \texttt{Matchete} and
calculated their matching. The results are summarised in table
\ref{tab:CXY_2VLF}. First, note that while for all models there are
two independent Yukawa couplings $y_{L/R}$, we always assume in the
matching $y_{L}=0$ and consider only $y_R$, since for $y_L=y_R$ all
NTGCs vanish identically, as discussed in the case of the prototype
model. As in table \ref{tab:ciHL}, we again quote directly
$\tilde{c}_{DAB}$. It is worth mentioning that for every model shown
here, the tensor couplings between $SU(2)$ multiplets are all
normalised to the square root of the product of the dimensions of the
representations. For example, the tensor defining the coupling of a
$\mathbf{4}$-plet, triplet and doublet is normalised to 
$\sqrt{4 \times 3 \times 2}$. 

We also need to mention briefly that for any multiplet other than the
six minimal fermionic multiplets shown in table \ref{tab:ciHL}, it is
not possible to write down a perturbative coupling that allows the
lightest state in that multiplet to decay to SM particles. Such \linebreak states
therefore would be absolutely stable.\footnote{The loophole to this
  argument is that these particles could still decay via
  higher-dimensional non-renormalizable operators, if no symmetry
  forbids it.} Experimental searches for stable charged relics
essentially exclude the mass range $M \sim [1, 10^5]$ GeV, see for
example
\cite{ParticleDataGroup:2020ssz,Hemmick:1989ns,Kudo:2001ie,Taoso:2007qk}.
The simplest ``solution'' to this problem is, of course, to consider
the models shown in table \ref{tab:CXY_2VLF} as series: For example, a
model, such as MDS4 with $(F_{1,2,-5/2},F_{1,1,-2})$ should also
contain $F_{1,2,-3/2}$, which then can decay to SM fields. This means
that one would need to add fields with smaller representations and/or
lower hypercharges for some models with large multiplets and/or
hypercharges.

\begin{table}[th!]
\centering
\resizebox{!}{.335\paperheight}{
\begin{tabular}{|l|>{$}l<{$}|>{$}c<{$}|>{$}c<{$}|>{$}c<{$}|>{$}c<{$}|}
\hline
Model & \text{Particles} & \tilde{c}_{DB\tilde{B}} & \tilde{c}_{DW\tilde{B}} = \tilde{c}_{DB\tilde{W}}   & \tilde{c}_{DW\tilde{W}} \\
\hline
\hline
MDS1 & (L_H, E_H) & \frac{23}{960} & -\frac{7}{480}   & \frac{1}{320} \\
\hline
MDS2 & (F_{1,2,-\frac{3}{2}}, F_{1,1,-1}) & -\frac{21}{320} & -\frac{13}{480}   & -\frac{1}{320} \\
\hline
MDS3 & (F_{1,2,-\frac{3}{2}}, F_{1,1,-2}) & \frac{41}{320} & -\frac{17}{480}  & \frac{1}{320} \\
\hline
MDS4 & (F_{1,2,-\frac{5}{2}}, F_{1,1,-2}) & -\frac{203}{960} & -\frac{23}{480}  & -\frac{1}{320} \\
\hline
MDS5 & (F_{1,2,-\frac{5}{2}}, F_{1,1,-3}) & \frac{101}{320} & -\frac{9}{160}   & \frac{1}{320} \\
\hline
MDS6 & (F_{1,2,-\frac{7}{2}}, F_{1,1,-3}) & -\frac{141}{320} & -\frac{11}{160}  & -\frac{1}{320} \\
\hline
MDS7 & (F_{1,2,-\frac{7}{2}}, F_{1,1,-4}) & \frac{563}{960} & -\frac{37}{480}   & \frac{1}{320} \\
\hline
\hline
\hline
MTD1 & (F_{1,3,0}, F_{1,2,-\frac{1}{2}}) & -\frac{\sqrt{3}}{320} & \frac{11}{480 \sqrt{3}}   & -\frac{49}{960 \sqrt{3}} \\
\hline
MTD2 & (F_{1,3,-1}, F_{1,2,-\frac{1}{2}}) & \frac{23}{320 \sqrt{3}} & \frac{13}{160 \sqrt{3}}  & \frac{49}{960 \sqrt{3}} \\
\hline
MTD3 & (F_{1,3,-1}, F_{1,2,-\frac{3}{2}}) & -\frac{21 \sqrt{3}}{320} & \frac{61}{480 \sqrt{3}}   & -\frac{49}{960 \sqrt{3}} \\
\hline
MTD4 & (F_{1,3,-2}, F_{1,2,-\frac{3}{2}}) & \frac{41 \sqrt{3}}{320} & \frac{89}{480 \sqrt{3}}  & \frac{49}{960 \sqrt{3}} \\
\hline
MTD5 & (F_{1,3,-2}, F_{1,2,-\frac{5}{2}}) & -\frac{203}{320 \sqrt{3}} & \frac{37}{160 \sqrt{3}}   & -\frac{49}{960 \sqrt{3}} \\
\hline
\hline
\hline
MQT1 & (F_{1,4,-\frac{1}{2}}, F_{1,3,0}) & -\frac{\sqrt{\frac{3}{2}}}{160} & -\frac{19}{240 \sqrt{6}} &   -\frac{109}{480 \sqrt{6}} \\
\hline
MQT2 & (F_{1,4,-\frac{1}{2}}, F_{1,3,-1}) & \frac{23}{160 \sqrt{6}} & -\frac{17}{80 \sqrt{6}} & \frac{109}{480 \sqrt{6}} \\
\hline
MQT3 & (F_{1,4,-\frac{3}{2}}, F_{1,3,-1})  & -\frac{21 \sqrt{\frac{3}{2}}}{160} & -\frac{89}{240 \sqrt{6}}   & -\frac{109}{480 \sqrt{6}} \\
\hline
MQT4 & (F_{1,4,-\frac{3}{2}}, F_{1,3,-2}) & \frac{41 \sqrt{\frac{3}{2}}}{160} & -\frac{121}{240 \sqrt{6}}   & \frac{109}{480 \sqrt{6}} \\
\hline
MQT5 & (F_{1,4,-\frac{5}{2}}, F_{1,3,-2}) & -\frac{203}{160 \sqrt{6}} & -\frac{53}{80 \sqrt{6}}  & -\frac{109}{480 \sqrt{6}} \\
\hline
\hline
\hline
MQQ1 & (F_{1,5,0}, F_{1,4,-\frac{1}{2}}) & -\frac{1}{32 \sqrt{10}} & \frac{7}{48 \sqrt{10}}   & -\frac{21}{32 \sqrt{10}} \\
\hline
MQQ2 & (F_{1,5,-1}, F_{1,4,-\frac{1}{2}}) & \frac{23}{96 \sqrt{10}} & \frac{23}{48 \sqrt{10}}   & \frac{21}{32 \sqrt{10}} \\
\hline
MQQ3 & (F_{1,5,-1}, F_{1,4,-\frac{3}{2}}) & -\frac{21}{32 \sqrt{10}} & \frac{37}{48 \sqrt{10}}  & -\frac{21}{32 \sqrt{10}} \\
\hline   
MQQ4 & (F_{1,5,-2}, F_{1,4,-\frac{3}{2}}) & \frac{41}{32 \sqrt{10}} & \frac{53}{48 \sqrt{10}}   & \frac{21}{32 \sqrt{10}} \\
\hline
MQQ5 & (F_{1,5,-2}, F_{1,4,-\frac{5}{2}}) & -\frac{203}{96 \sqrt{10}}  & \frac{67}{48
\sqrt{10}} & -\frac{21}{32 \sqrt{10}} \\
\hline
\end{tabular}
}

\caption{
Model name, particle content and matching for the $\tilde{c}_{DAB}$ 
for different models with two BSM VLFs. In all cases we assume only one 
Yukawa coupling to be non-zero, see also text.}
\label{tab:CXY_2VLF}
\end{table}

It is actually possible to derive an analytic formula for
$\tilde{c}_{DAB}$ for the models given in table
\ref{tab:CXY_2VLF}. The coefficients are given by:
\begin{eqnarray}\label{eq:cBtBA}
\centering
\tilde{c}_{DB\tilde{B}} &=&  \frac{1}{160} (-1)^{(\mathbf{r_1} \bmod 2)}  \text{sgn}\!\left(y_2^2-y_1^2\right) \sqrt{2\mathbf{r_1}\mathbf{r_2} } \left( y_1^2 + y_2^2 +\frac{4}{3} y_2 y_1 \right)  \, ,
\\ \nonumber
\tilde{c}_{DW\tilde{W}} &=& \frac{1}{160} (-1)^{(\mathbf{r_1} \bmod 2)}  \text{sgn}\!\left(y_2^2-y_1^2\right) \sqrt{2\mathbf{r_1}\mathbf{r_2} }\,  \frac{1}{12} \left[(\mathbf{r_1}^2-1)+(\mathbf{r_2}^2-1)+\frac{4}{3} \left(\mathbf{r_1} \mathbf{r_2}-2\right)\right] \, ,
\\ \nonumber
\tilde{c}_{DW\tilde{B}} &=& \frac{1}{48} (-1)^{(\mathbf{r_1} \bmod 2)} \sqrt{2\mathbf{r_1}\mathbf{r_2} } \,\frac{1}{12}  \left[\left(y_1+y_2\right) \left(\mathbf{r_1}+\mathbf{r_2}\right)+\frac{3}{5} \left(y_1-y_2\right) \right]\, ,
\\ \nonumber
\tilde{c}_{DB\tilde{W}} &=& \tilde{c}_{DW\tilde{B}} \, .
\end{eqnarray}
Here, the input is the $SU(2)_L$ representations and hypercharges for
each pair of fermions in table \ref{tab:CXY_2VLF} as $(\mathbf{r_1},y_1)$ and $(\mathbf{r_2},y_2)$, respectively.

These equations allow to calculate the vertices $f_5^V$ and $h_3^V$
also for models with larger representations and hypercharges than the
ones given in table \ref{tab:CXY_2VLF}.  Note that the equations are
valid only for models with the VLF with representation and hypercharge
differences fixed by their Yukawa coupling to the SM Higgs,
i.e. $|\mathbf{r_2}-\mathbf{r_1}|=1$ and $|y_2-y_1|=1/2$.

Several clear patterns emerge when one compares the different
coefficients shown in table \ref{tab:CXY_2VLF}. Trivially and as
expected, the coefficients increase with increasing multiplet size and
hypercharge. Comparing different models, the largest coefficient
$\tilde{c}_{DB\tilde{B}}$ is found for MQQ5. This model has a
coefficient larger than the prototype model MDS1 by a factor $r =
203/23 \sqrt{10} \simeq 28$. However, note that even this large ratio
will lead to a change in the scale $\Lambda$ of the operator only by a
factor of roughly $2.3$.

\begin{figure}[t!]
    \centering
    \includegraphics[scale=0.9]{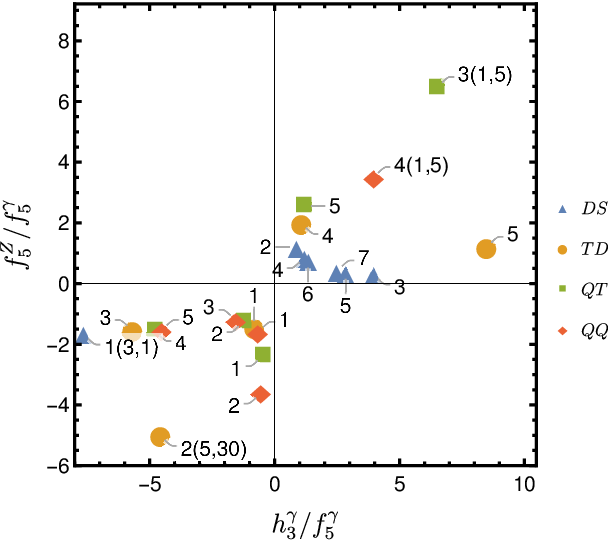}
    \caption{Ratios $h_3^{\gamma}/f_5^{\gamma}$ versus
      $f_5^Z/f_5^{\gamma}$ for the different two-VLF models of table
      \ref{tab:CXY_2VLF}.  Model classes are identified by symbols,
      the numbers identify the models in the corresponding
      sub-class. Brackets indicate that the value for the point shown
      has to be scaled by factors ($x,y$) to obtain the true
      ratios. This scaling was done only for presentation purposes.}
    \label{fig:NTGCrats}
\end{figure}

More interesting is the fact that the relative ratios of the different 
$\tilde{c}_{DAB}$ that one finds for the different models lead to 
different predictions for the ratios of measurable couplings 
$h_3^V$, $f_5^V$. Note, in all our models $h_3^Z = -f_5^{\gamma}$, 
i.e. from the remaining three free couplings we can form two 
independent ratios, say $h_3^{\gamma}/f_5^{\gamma}$  and $f_5^Z/f_5^{\gamma}$. 
We plot these ratios for all models with two VLFs in figure \ref{fig:NTGCrats}.
Note that by plotting ratios of form factors both the overall 
scale $\Lambda$ of the operators involved and the Yukawa 
couplings cancel, thus these ratios are strict predictions of the
models discussed here.

One notices that all models have either both ratios positive or both
ratios negetive. Moreover, very different ratios can occur in
different models, thus if these form factors ever could be measured,
their ratios would be an interesting model discriminator, as this plot
clearly demonstrates.

\begin{figure}[t!]
    \centering
    \includegraphics[scale=0.69]{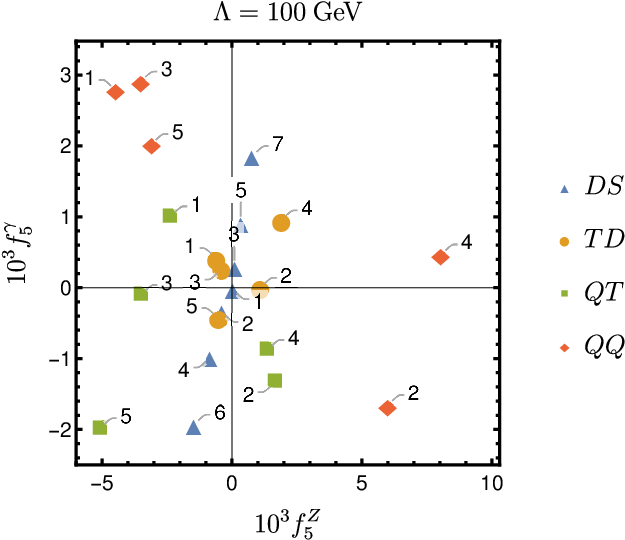}\hskip3mm
    \includegraphics[scale=0.69]{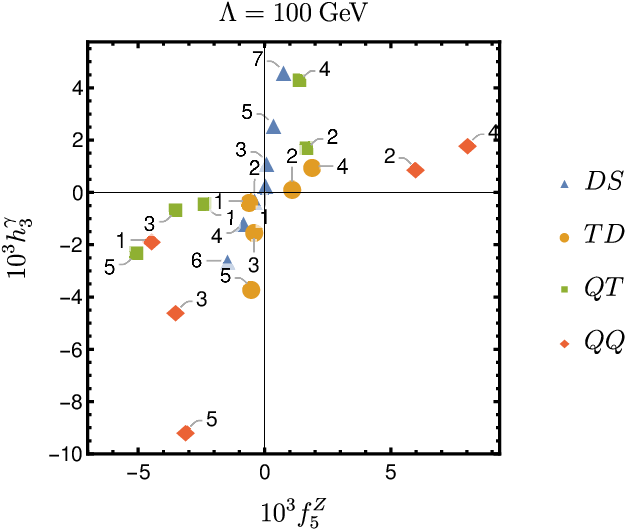}
    \caption{$f_5^{Z}$ versus $f_5^{\gamma}$ (left) and $h_3^{\gamma}$
      (right) for the different two-VLF models of table
      \ref{tab:CXY_2VLF} for the particular choice of $\Lambda=100$
      GeV.  Model classes are identified by symbols, the numbers
      identify the models in the corresponding sub-class. }
    \label{fig:f5}
\end{figure}

In figure \ref{fig:f5} we plot the absolute values of the form factors
$f_5^{Z}$ versus $f_5^{\gamma}$ and $h_3^{\gamma}$ for all models of
table \ref{tab:CXY_2VLF}.  This plot assumes a scale of $\Lambda=100$
GeV and sets all Yukawa couplings exactly to one. The scale of
$\Lambda=100$ GeV has been chosen such that the coefficients of the
models are roughly order ${\cal O}(10^{-3})$. The largest value for
the form factors are found for the models QQ4 and QQ5.  It is also
striking that for the DS models $f_5^{Z}$ is always small, but 
a ``large'' $h_3^\gamma$ can be generated simply by choosing large 
values for the hypercharge for the VLFs.

\begin{figure}[t!]
    \centering
    \includegraphics[scale=0.65]{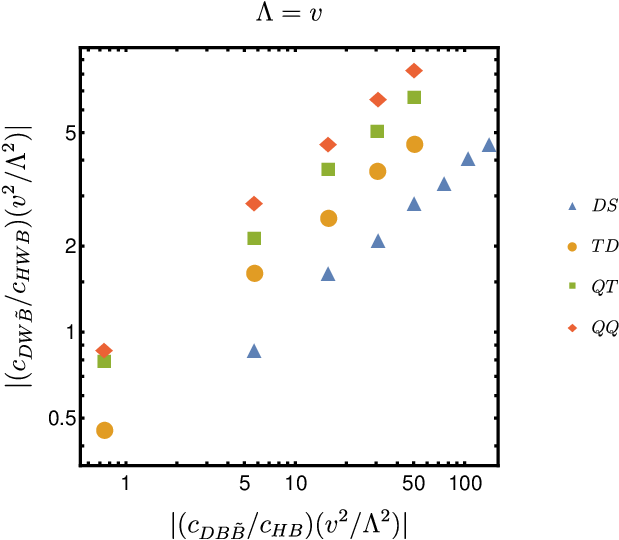}
    \hskip5mm
    \includegraphics[scale=0.65]{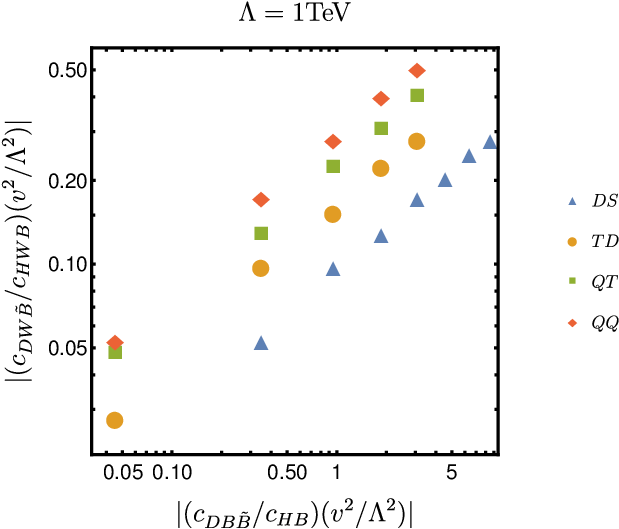}
    \caption{Ratios of coefficients for the $d=8$ operators
      $c_{DB\tilde{B}}/\Lambda^4$ ($c_{DW\tilde{B}}/\Lambda^4$)
      divided by the coefficient of the $d=6$ operators
      $c_{HB}/\Lambda^2$ ($c_{HWB}/\Lambda^2$) for the different
      models listed in table \ref{tab:CXY_2VLF}. To get a
      dimensionless ratio, we scale by $v^2$ and show ratios for two
      examples of $\Lambda$. The absolute values of the ratios change
      for different choices of $\Lambda$, but the patterns remain
      unchanged.  The models are always ordered such that the models
      with the smallest number, see table \ref{tab:ciHL}, have the
      smallest ratios.}
    \label{fig:Rat68}
\end{figure}

Finally, we need to discuss briefly also bosonic $d=6$
operators. Models that generate the $d=8$ NTGC operators will
necessarily also generate the following four bosonic $d=6$ operators:
\footnote{There are also the $d=6$ CPV operator variants
  ${\cal O}_{H\tilde{X}}$. These are less constrained experimentally, so we
  do not give their explicit matching for all models. We only note in
  passing that in order to generate the $d=6$ CPV operators, two
  Yukawa couplings $y_L$ and $y_R$ are needed and it is required that
  $y_L \ne y_R$.  For the prototype model this is discussed in
  section \ref{sect:pheno}.}
\begin{eqnarray}\label{eq:OHB}
{\cal O}_{HB} &=& \frac{c_{HB}}{\Lambda^2} H^{\dagger}HB_{\mu\nu}B^{\mu\nu},
\\ \label{eq:OHWB}
{\cal O}_{HWB} &=& \frac{c_{HWB}}{\Lambda^2}H^{\dagger}HW_{\mu\nu}B^{\mu\nu},
\\ \label{eq:OHW}
{\cal O}_{HW} &=& \frac{c_{HW}}{\Lambda^2}H^{\dagger}HW_{\mu\nu}W^{\mu\nu},
\end{eqnarray}
and
\begin{eqnarray}\label{eq:OW}
{\cal O}_{3W} &=& \frac{c_{3W}}{\Lambda^2}W_{\mu}^{\nu}W_{\nu}^{\rho}W^{\mu}_{\rho}
\end{eqnarray}
Table \ref{tab:ci_dim6} gives the matching of the Wilson coefficients
for these operators for our different model classes with two heavy
fermions. Table \ref{tab:HL_dim6} shows the matching for models with
only one heavy fermion, for completeness.  Interestingly, the
coefficients for these operators do only depend on the combinations of
$SU(2)_L$ representations in the models and not on the choice of
hypercharge, i.e.\ all models within one class (DS, TD, QT or QQ) have
the same coefficient.  The reason for this is that in the matching of
${\cal O}_{HB}$, ${\cal O}_{HWB}$, and ${\cal O}_{HW}$ the difference
of the hypercharges appears (due to the coupling to the Higgs) and
this difference is always $1/2$ in all models. As the table also
shows, the matching of the first three operators are proportional to
the Yukawa coupling squared, exactly the same as the $d=8$ NTGC
operators. Taking ratios, such as $c_{DB\tilde{B}}/c_{HB}$, will
therefore cancel the dependence on the unknown Yukawa couplings.
 
\begin{table}[h]
\centering
\begin{tabular}{|l|c|c|c|c|}
\hline
 & $c_{HB}$ & $c_{HWB}$ & $c_{HW}$ & $c_{3W}$  \\
\hline
MDS & $-\frac{1}{240} g_Y^2 |y_i|^2$ & $-\frac{1}{60} g_L g_Y |y_i|^2$ & $-\frac{1}{240} g_L^2 |y_i|^2$ & $-\frac{1}{180} g_L^3$ \\
\hline
MTD & $-\frac{1}{80\sqrt{3}} g_Y^2 |y_i|^2$ & $-\frac{1}{20\sqrt{3}} g_L g_Y |y_i|^2$ & $-\frac{1}{80\sqrt{3}} g_L^2 |y_i|^2$ & $-\frac{1}{36} g_L^3$ \\
\hline
MQT & $-\frac{1}{40\sqrt{6}}g_Y^2 |y_i|^2$ & $-\frac{1}{10 \sqrt{6}} g_L g_Y |y_i|^2$ & $-\frac{1}{40 \sqrt{6}} g_L^2 |y_i|^2$ & $-\frac{7}{90} g_L^3$ \\ 
\hline
MQQ & $-\frac{1}{24 \sqrt{10}} g_Y^2 |y_i|^2$ & $-\frac{1}{6 \sqrt{10}} g_L g_Y |y_i|^2$ & $-\frac{1}{24 \sqrt{10}} g_L^2 |y_i|^2$ & $-\frac{1}{6} g_L^3$ \\
\hline
\end{tabular}
\caption{Wilson coefficients for bosonic dimension-6 operators for the two
  particle models listed in table \ref{tab:CXY_2VLF}. A common factor
  $(16\pi^2)^{-1}$ has been supressed in all cases.}
\label{tab:ci_dim6}
\end{table}

\begin{table}[h]
\centering
\begin{tabular}{|l|c|c|c|c|}
\hline
 & $c_{HB}$ & $c_{HWB}$ & $c_{HW}$ & $c_{3W}$ \\
\hline
Type-I & 0 & 0 & 0 & 0 \\
\hline
Type-III & $\frac{1}{8 \sqrt{3}} g_Y^2 |y_{l\Sigma}|^2$ & $-\frac{1}{2\sqrt{3}} g_L g_Y |y_{l\Sigma}|^2$ & $\frac{7}{24 \sqrt{3}} g_L^2 |y_{l\Sigma}|^2$ & $-\frac{1}{45} g_L^3$ \\
\hline 
MLH & $\frac{1}{4} g_Y^2 |y_{Le}|^2$ & $-\frac{1}{6} g_L g_Y |y_{Le}|^2$ & $0$ & $ -\frac{1}{180}g_L^3$\\
\hline
MEH & $\frac{1}{8} g_Y^2 |y_{lE}|^2$ & $-\frac{1}{3} g_L g_Y |y_{lE}|^2$ & $\frac{1}{24} g_L^2 |y_{lE}|^2$ & 0 \\
\hline
MFD & $\frac{5}{12} g_Y^2 |y_{F32}|^2$ & $\frac{1}{6} g_L g_Y |y_{F32}|^2$ & $0$ & $-\frac{1}{180} g_L^3$\\
\hline
MFT & $\frac{3}{8 \sqrt{3}} g_Y^2 |y_{l\Psi}|^2$ & $\frac{2}{3\sqrt{3}} g_L g_Y |y_{l\Psi}|^2$ & $\frac{7}{24\sqrt{3}} g_L^2 |y_{l\Psi}|^2$ & $-\frac{1}{45} g_L^3$ \\
\hline
\end{tabular}
\caption{Wilson coefficients for bosonic $d=6$ operators for one-heavy
  particle models given in table \ref{tab:ciHL}. All coefficients must
  be multiplied by a common 1-loop factor of $(16\pi^2)^{-1}$.}
\label{tab:HL_dim6}
\end{table}

Figure \ref{fig:Rat68} shows the ratios of $c_{DB\tilde{B}}$, the
Wilson coefficients for the $d=8$ operator ${\cal O}_{DB\tilde{B}}$,
divided by $c_{HB}$ versus the ratio $c_{DW\tilde{B}}/c_{HWB}$. Since
the $d=6$ ($d=8$) operator scales as $1/\Lambda^2$ ($1/\Lambda^4$), we
multiply the ratio by $v^2$ to get a dimensionless number and show two
plots with different choices of $\Lambda$. The different sub-types of
models are distinguished by different symbols. In each sub-class the
two ratios always increase monotonously with hypercharge, compare to
table \ref{tab:CXY_2VLF}, thus models ``1'' are the left-most points
in each class in this plot. For $\Lambda$ as low as $\Lambda = 100 $
GeV for many of the models the ratios are larger than one, implying
that the $d=8$ operators are more important than the $d=6$
operators. Even for $\Lambda = 1 $ TeV ratios of order one (and larger)
can be found for the models with the largest hypercharges.  It is
interesting to note also that in this parameter plane all models lie
on different coordinates and thus, in principle, could be
distinguished experimentally, if all four Wilson coefficients could
ever be measured.

We also mention that the ratio of $c_{DW\tilde{W}}/c_{HW}$ is simply
given by $c_{DW\tilde{W}}/c_{HW} = f/\Lambda^2$, where $f$ is a
constant within each model class and $|f|=(3/4,49/12,109/12,63/4)$ for
(DS, TD, QT, QQ), respectively. (The sign of $f$ is alternating within
a model class as can be read off from table \ref{tab:CXY_2VLF}.) Thus,
measuring this ratio would fix the scale $\Lambda$ uniquely within
each model class.

Finally we mention that all models will unavoidably generate 
also ${\cal O}_{3W}$. This operator only depends on the scale $\Lambda$, 
but does not involve the (unknown) Yukawa couplings. The  relative 
importance of ${\cal O}_{3W}$ and the other $d=6$ operators can therefore 
not be predicted. Since  ${\cal O}_{3W}$ is a 1-loop generated 
operator, however, the coefficients are also rather small, 
but increase rapidly with the size of the $SU(2)_L$ representations 
in the model.


\section{Phenomenology\label{sect:pheno}}

In the previous section we have discussed UV models, which lead
to non-zero neutral triple gauge boson couplings (NTGCs). Those UV
scenarios also produce correlated signatures, stemming from
their contribution to dimension-six operators and through direct
searches. In this section we will discuss the current and future LHC
reach from all these signatures.

\subsection{Dimension-six contributions}

Before we move to discuss the contributions of dimension-eight
operators to NTGCs, we examine the correlated SMEFT contributions to
dimension-six operators. The energy dependence of these operators is
milder than the $d=8$ contributions, as we discussed in
section~\ref{sect:models} and illustrated in figure \ref{fig:Rat68}.  
Nevertheless, the $d=6$ contributions need to
be evaluated alongside $d=8$. We will proceed to discuss this issue within scenarios with vector-like fermions, but general arguments can also be found in several works, see e.g.~\cite{Arzt:1994gp,Buchalla:2022vjp}.

Let us start with the simplest possible extension of the SM with
vector-like fermions, a scenario we presented in
section~\ref{sect:proto}. As explained in that section, to generate the
NTGC operators at least one Yukawa coupling is needed. The discussion
on the prototype model can be generalised to the more complex
benchmarks we will consider later.

For these fields we can write down four different Yukawa couplings,
see \eqref{eq:LagLE}. The Yukawas $y_{Le}$ and $y_{lE}$ connect the
heavy fermions to their SM counter-parts, while for $y_R$ and $y_L$
only the heavy fermions are involved. Thus, one can expect that there
are important tree-level constraints on $y_{Le}$ and $y_{lE}$, while
$y_{L/R}$ will show up only (i) multiplied via either $y_{Le}$ or
$y_{lE}$ in a tree-level generated operator or (ii) without $y_{Le}$
or $y_{lE}$ at 1-loop level. This simple argument shows that
$y_{Le}$ and $y_{lE}$ will be much more constrained than
$y_{L/R}$. We now discuss in turn these two possibilities.

{\bf Tree-level $d=6$ operators:} The prototype model generates only
four $d=6$ SMEFT operators at tree-level.  These are the operators
${\cal O}_{Hl}^{(1)}$, ${\cal O}_{Hl}^{(3)}$, ${\cal O}_{He}$ and
${\cal O}_{eH}$. The matching for these operators is given by:
\begin{eqnarray}\label{eq:mtchHl1}
c_{Hl^{(1)}} & = & -\frac{1}{4} |y_{lE}|^2, \\ \label{eq:mtchHl3}
c_{Hl^{(3)}} & = & -\frac{1}{4} |y_{lE}|^2, \\ \label{eq:mtchH3}
c_{He} & = & \frac{1}{2} |y_{Le}|^2, \\ \label{eq:mtch3H}
c_{eH} & = &   - y_L^{\dagger}y_{lE}y_{Le}. 
\end{eqnarray}
According to the global fit analysis done in \cite{Ellis:2020unq}, 
the individual (marginalised) constraints on the Wilson coefficients 
for these operators are:
\begin{eqnarray}\label{eq:LimHl1}
  c_{Hl^{(1)}} & \to & \frac{\Lambda}{\sqrt{c_{Hl^{(1)}}}} \ge 11 (1.8) \hskip1mm {\rm TeV}
  \\ \label{eq:LimHl3}
c_{Hl^{(3)}} & \to & \frac{\Lambda}{\sqrt{c_{Hl^{(3)}}}} \ge 12 (3.3) \hskip1mm {\rm TeV} 
  \\ \label{eq:LimHe}
c_{He} & \to &   \frac{\Lambda}{\sqrt{c_{He}}} \ge 9.6 (1.3) \hskip1mm {\rm TeV} 
\end{eqnarray}
This allows us to estimate the upper limit on the couplings:
\begin{eqnarray}\label{eq:LimHl1a}
|y_{lE}| \le 0.18 (1.11) 
  \\ \label{eq:LimHl3a}
|y_{lE}| \le 0.167 (0.61) 
  \\ \label{eq:LimHea}
|y_{Le}| \le 0.147 (1.09) 
\end{eqnarray}
for $\Lambda=1$ TeV. Note, though, the energy dependence of these
limits, which would be one order of magnitude more stringent for
$\Lambda=100$ GeV. Recall, that LEP essentially rules out {\em any
  charged} fermion with a mass below roughly 100 GeV. Limits on $L_H$
and $E_H$ from LHC are more stringent, but may contain loopholes.

To summarise, the limits on the couplings $y_{Le}$ and $y_{lE}$ from
these operators compensate by far the larger coefficients that one
finds for the NTGC operators in the matching of heavy-light loops
relative to heavy-heavy loops. This motivates looking only at
$y_{L/R}$ when matching NTGCs.

{\bf Matching at 1-loop $d=6$ operators:} Many operators appear at
1-loop order already at $d=6$. Especially, in the on-shell basis many
4F operators appear which are not proportional to Yukawa couplings on
the flavour-diagonal.  More important, however, is the fact that the
model also produces bosonic operators. An example is ${\cal O}_{3W}$,
which has the matching:
\begin{equation}\label{eq:O3W}
  c_{3W}= - \frac{1}{16 \pi^2}\frac{g_L^3}{180}
       \hskip5mm \simeq 9 \times 10^{-6} 
\end{equation}
In the global fit in \cite{Ellis:2020unq} the authors did not
consider this operator due to its inherent loop suppression. On the
other hand, the following six operators are ``unavoidable'' operators,
since they depend on {\em the same parameters as the NTGC
  operators}. These operators are: ${\cal O}_{HB}$, ${\cal O}_{HW}$,
${\cal O}_{HWB}$, ${\cal O}_{H\tilde{B}}$, ${\cal O}_{H\tilde{W}}$ and
${\cal O}_{H\tilde{W}B}$, see
\eqref{eq:OHB}-\eqref{eq:OW}. The CPC operators appear
already with only one non-zero Yukawa coupling, the CPV
operators need both $y_L$ {\em and} $y_R$.

The matching at 1-loop order, performed with \texttt{Matchete} \cite{Fuentes-Martin:2022jrf}, is found to be:
\begin{eqnarray}\label{eq:mtchHB}
  c_{HB} & = & -\frac{1}{16 \pi^2}\frac{g_Y^2}{240}
  \left[y_R^{\dagger}(46 y_L + y_R) + y_L^{\dagger}(46 y_R + y_L)\right] 
  \\ \label{eq:mtchHWB}
  c_{HWB} & = & -\frac{1}{16 \pi^2}\frac{g_Yg_L}{60}
  \left[y_R^{\dagger}(-14 y_L + y_R) + y_L^{\dagger}(-14 y_R + y_L)\right] 
  \\ \label{eq:mtchHW}
  c_{HW} & = & -\frac{1}{16 \pi^2}\frac{g_L^2}{240}
  \left[y_R^{\dagger}(6 y_L + y_R) + y_L^{\dagger}(6 y_R + y_L)\right] 
\end{eqnarray}
for the CPC operators, while for the CPV operators one finds:
\begin{eqnarray}\label{eq:mtchHtB}
  c_{H\tilde{B}} & = & \frac{1}{16 \pi^2}\,i\frac{7 g_Y^2}{48}
  \left(y_L^{\dagger}y_R - y_R^{\dagger}y_L\right) 
  \\ \label{eq:mtchHtWB}
  c_{H\tilde{W}B} & = & -\frac{1}{16 \pi^2}\, i\frac{g_Yg_L}{6}
  \left(y_L^{\dagger}y_R - y_R^{\dagger}y_L\right) 
  \\ \label{eq:mtchHtW}
  c_{H\tilde{W}} & = & \frac{1}{16 \pi^2}\, i\frac{g_L^2}{48}
  \left(y_L^{\dagger}y_R - y_R^{\dagger}y_L\right) 
\end{eqnarray}
A few comments are in order. First of all, all CPV operators
vanish identically in the limit $y_L = y_R$ and in the
cases when either $y_L$ or $y_R$ vanish. This is not surprising,
since CPV requires a non-zero phase and in both these limits,
there is only one possible phase in the remaining Yukawa coupling
and an overall phase can not have any physical consequences.

Second, if we approximate $y_R^{\dagger}(46 y_L + y_R) \simeq 46
y_R^{\dagger} y_L$, $c_{HB}$ would vanish for the choice: $y_L=y
\times \exp{(i \phi_L)}$, $y_R=y \times \exp{(i \phi_R)}$ and $\phi_R
= \phi_L + \pi/2$. More exactly, $c_{HB}$ vanishes for $\phi_R \simeq
\phi_L + 1.01384 \frac{\pi}{2}$. Of course, this is only one
particular and fine-tuned point in parameter space and it seems
impossible to cancel all three $c_{HB}$, $c_{HW}$ and $c_{HWB}$ at the
same time by such a fine-tuning.  However, it is worth mentioning that
such a fine-tuning at the same time would (nearly) maximise the CPV
operators. The CPC and CPV Higgs-gauge operators could be
distinguished through a set of experimental measurements, see
e.g. \cite{Ferreira:2016jea,Cirigliano:2019vfc,Biekotter:2021int}
for studies on the CPC vs CPV operators.

According to table 6 of \cite{Ellis:2020unq}, there are
the following individual (marginalised) limits on the
Wilson coefficients of the CPC operators:
\begin{eqnarray}\label{eq:LimHB}
  c_{HB} & \to & \frac{\Lambda}{\sqrt{c_{HB}}} \ge 17 (1.4) \hskip2mm {\rm TeV}
  \\ \label{eq:LimHWB}
  c_{HWB} & \to & \frac{\Lambda}{\sqrt{c_{HWB}}} \ge 19 (2.5) \hskip2mm {\rm TeV}
  \\ \label{eq:LimHW}
  c_{HW} & \to  & \frac{\Lambda}{\sqrt{c_{HW}}} \ge 11 (1.4) \hskip2mm {\rm TeV}
\end{eqnarray}
A model that generates all three of them, as in our benchmark
scenarios, may not be able to escape the stronger constraint, although
one should re-do the global fit with those operator correlations to
verify this statement.

Due to the large relative factor (46, in case of $c_{HB}$) between
the cases with only $y_{L}$ or $y_{R}$ and the case where both
couplings are non-zero, limits strongly depend on the scenario.
For the simplest benchmarks, and using the strongest bound from the three limits in Eq.\ref{eq:LimHB}, one finds:
\begin{eqnarray}\label{eq:LimHB2}
  \Lambda \ge 31 (2.6) \hskip2mm {\rm GeV} & \hskip5mm & 
  (y_L=1,y_R=0) \hskip1mm {\rm or } \hskip1mm (y_L=0,y_R=1)
    \\ \label{eq:LimHB2a}
  \Lambda \ge 301 (24.8) \hskip2mm {\rm GeV} & \hskip5mm & y_L=y_R=1
\end{eqnarray}
The individual $y_L=y_R=1$ limit is much stronger than what the NTGCs
will give us, but all others are actually weaker than the ones
obtained from the NTGCs.

To conclude, in this section we have discussed the limits coming from
the contribution of fermionic models to $d=6$ operators. We have found
that the sets of models which can be probed with NTGCs produce
contributions to $d=6$ operators at 1-loop. We have discussed the
matching of the prototype model (MDS1) benchmark, the current limits from global fits to
$d=6$ operators and its interplay with NTGCs. The discussion can be
extended to the other benchmarks, using the matching to $d=6$
operators provided in tables~\ref{tab:ci_dim6} and~\ref{tab:HL_dim6}
for models with two and one VLFs, respectively.

\subsection{Neutral Triple Gauge Couplings (NTGCs)}~\label{sec:NTGCpheno}

\begin{figure}[t!]
\begin{center}
\includegraphics[scale=1.]{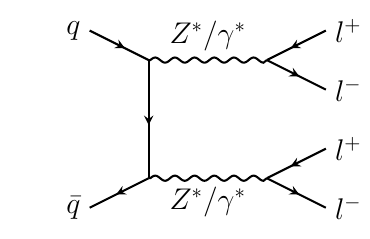}
\includegraphics[scale=1.]{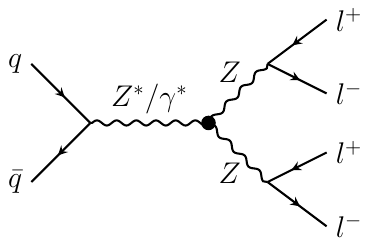}
\caption{Diagrams contributing to the four-lepton final state: a
  typical standard model contribution ({\it left}) and the leading
  contribution from dimension-eight operators ({\it right}).}
\label{fig:diagramsZZ}
\end{center}
\end{figure}

We now move to the search for neutral triple gauge couplings and how
they relate to our benchmarks. Starting with the current constraints
in the next section, mostly from the four-lepton $ZZ$ final state, we
will estimate the projected sensitivity in the high-luminosity phase
of the LHC.

\subsubsection{Four-lepton (4L) analysis}~\label{sec:4L}

With the Run2 dataset, ATLAS~\cite{ATLAS:2021kog} and
CMS~\cite{CMS:2020gtj} have measured differential distributions of the
production of two $Z$ bosons, with the $Z$ decaying into two leptons,
$\ell=e^\pm$ or $\mu^\pm$. This four-lepton final state channel, which
we will represent by $4L$, is extremely clean and provides precious
information on possible NTGCs.\footnote{Additionally, one can consider
  final states with one photon and a $Z$ boson, although this channel
  provides weaker constraints on the NTGCs. See
  \cite{ATLAS:2019gey,CMS:2015wtk} for experimental analyses in
  the $\gamma+$2L final state.} The main SM background comes from
the t-channel contribution to a $ZZ$ final state, as shown in the left
side of figure \ref{fig:diagramsZZ}. New physics via NTGCs could
contribute to this final state through the diagram shown on the right
panel of figure \ref{fig:diagramsZZ}.

\begin{figure}[t!]
\begin{center}
\includegraphics[scale=0.7]{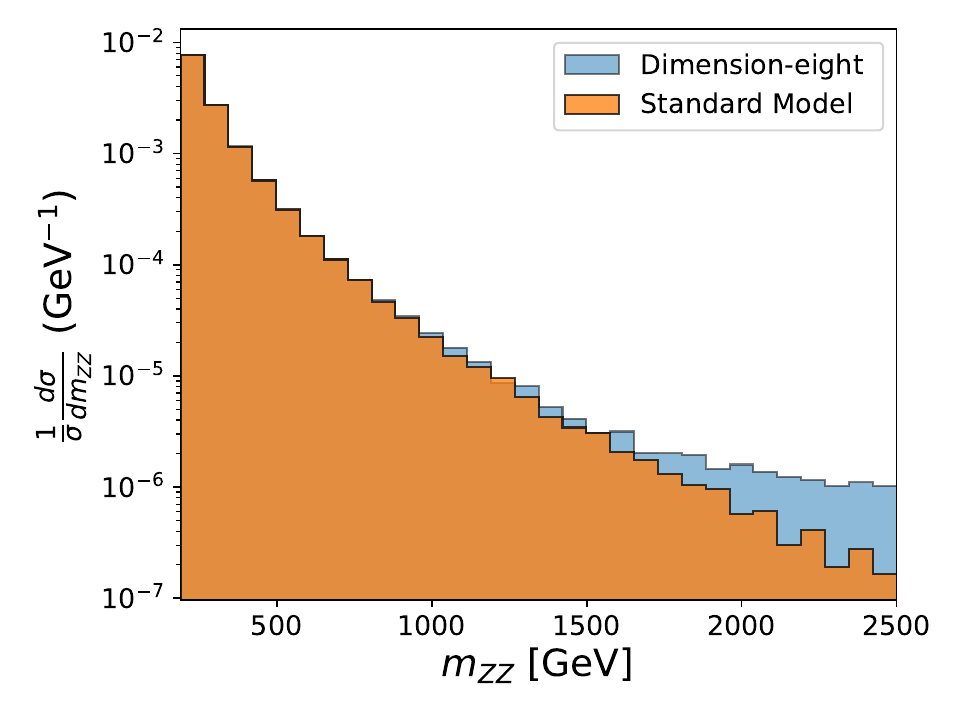}
\caption{Invariant mass distribution for the $ZZ$ final state at 13
  TeV. The orange and blue distributions correspond to switching off
  and on the dimension eight contributions. This example correspond to
  the choices $c_{D W \tilde W}$=1 and $\Lambda$ set to 1 TeV. }
\label{fig:distrib}
\end{center}
\end{figure}

In the region where the two $Z$ bosons are on-shell and boosted, the
t-channel SM 4L production is very suppressed, whereas the
dimension-eight contributions exhibit the kinematic growth discussed
in section~\ref{sect:operators}, see \eqref{eq.ntgcv1}. This is
illustrated in figure~\ref{fig:distrib}, where we plot the normalised
invariant mass distribution of the final state, $m_{ZZ}$. Even after
accounting for the PDF suppression, the growth of the distribution
with nonzero dimension-eight operators is visible, particularly at the
tail of the distribution.  Therefore, the last bin of the experimental
distributions will be particularly sensitive to the effect of NTGCs
and we will focus on it to set limits on our models.

We now summarise the two Run2 searches we have taken into account in
our analysis:

\vspace{.2cm}

{\bf $\bullet$ $Z Z \to $ 4L from ATLAS Run2 data:} The ATLAS
measurement in the region $m_{4\ell} \in [900, \, 1710]$ GeV is given
by a value $\frac{d \sigma}{d m_{4\ell}}= 1.6 \times 10^{-4}$ [fb/GeV]
with an error dominated so far by statistics, followed by
uncertainties in MC statistics and luminosity.  The bound in
cross-section from this bin is of the order of $1.6 \times 10^{-4}
\times 810 = $ 0.13 fb. From the experimental data and Monte-Carlo
distributions, we see that this value of the cross-section is
compatible with the SM (Powheg and Sherpa) predictions at 1$\sigma$,
which itself is of about 30\% of the measured central value. We can
use the compatibility of this measurement with the SM theoretical
expectation and the size of this error bar to set a limit on possible
NTGC contributions,
\begin{equation}
\sigma^{ATLAS, Run2}_{NTGC} (p p \to Z Z) \lesssim 0.08 \textrm{ fb at 95 \% C.L. in the bin } m_{4\ell} \in [900, \, 1710] \textrm{ GeV. }
\end{equation}

To set limits on the dimension-eight coefficients, we have simulated
the $m_{4\ell}$ distribution using our {\tt UFO}~\cite{Darme:2023jdn}
implementation in {\tt aMC@NLO}~\cite{Alwall:2014hca} and applied the
basic kinematic cuts $p_T^{\ell_1}> 20$ GeV, $\Delta
R_{\ell_1\ell_2}>$ 0.05.

{\bf $\bullet$ $Z Z \to 4 \ell$ from CMS Run2 data: } At 137
fb$^{-1}$, the measurement $1/\sigma d \sigma/ d m_{ZZ}$ (1/GeV) in
the bin $m_{ZZ}\in [800,1000]$ GeV is 1.79 $\times 10^{-5} \pm 7
\times 10^{-6}$~\cite{CMS:2020gtj}. We could use this measurement in
the same fashion we have done for ATLAS, but CMS provides an
additional sources of information. Besides giving the differential
information in bins with measured events, they also provide a
high-$m_{ZZ}$ bin in the range $>1300$ GeV,\footnote{Note that this
  information cannot be found in the {\tt HEPDATA} source in
  \url{https://doi.org/10.17182/hepdata.101183}, but in the published
  paper~\cite{CMS:2020gtj}, figure 7.} where no events have been
measured, but an estimate of the expected background is given
$N^{exp.}_{bg} = 2.8$ in the combined $4e$, $4\mu$ and $2e2\mu$
channels with 137 fb$^{-1}$ at 13 TeV.  When accounting for the
branching ratio of $ZZ\to 4 \ell$, this leads to a limit,
\begin{equation}
\sigma^{CMS, Run2}_{NTGC} (p p \to Z Z \to 4\ell) \lesssim 0.04  \textrm{ fb at 95 \% C.L. for } m_{4\ell} > 1300 \textrm{ GeV, }
\end{equation}
which can be compared with the theoretical prediction in this bin
obtained with the Monte-Carlo simulation.

Using both ATLAS and CMS results and comparing them with the
Monte-Carlo simulation in terms of the dimension-eight coefficients we
can build a distribution $\chi^2(c_i/\Lambda)$ where the $c_i$ are
possible values of $(c_{D B \tilde B}, \, c_{D W \tilde B}, \, c_{D B
  \tilde W}, \, c_{D W \tilde W})$. As explained in
section~\ref{sect:models}, the fermionic models produce a pattern, $\,
c_{D W \tilde B}=\, c_{D B \tilde W}$ and we can impose this condition
on the minimisation function, i.e. $\chi^2(c_{D B \tilde B}/\Lambda^4, \, c_{D W \tilde B}/\Lambda^4, \, c_{D W \tilde W}/\Lambda^4)$.

\begin{figure}[t!]
\begin{center}
\includegraphics[width=4in]{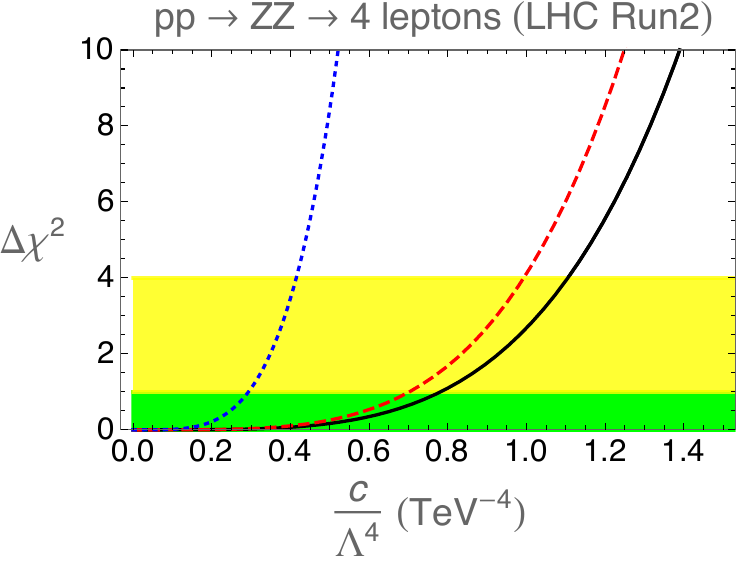}
\caption{Individual limits from the LHC Run2 data in the $ZZ$ final
  state. The lines represent the $\chi^2-\chi_{min}^2$ for $\, c_{D B
    \tilde B}$ (black-solid), $\, c_{D W \tilde B}=\, c_{D B \tilde
    W}$ (red-dashed) and $\, c_{D W \tilde W}$ (blue-dotted). The green and yellow areas correspond to 1- and 2-$\sigma$ for one degree of freedom.  }
\label{fig:ZZcurrent}
\end{center}
\end{figure}

In figure \ref{fig:ZZcurrent} we plot the $\chi^2$ distribution as a
function of individual dimension-eight parameters, i.e. setting two coefficients to zero every time and focusing on the limit from a single nonzero coefficient at a time. We find that the
current limit is,
\begin{eqnarray}\label{eq:limitscurrent}
\Lambda &>&  0.98 \,  c_{D B  \tilde B}^{1/4} \textrm{ TeV,} \nonumber \\
\Lambda &>&  1.00  \,   c_{D W  \tilde B}^{1/4} \textrm{ TeV, } \nonumber \\
\Lambda &>&  1.25  \,   c_{D W  \tilde W}^{1/4} \textrm{ TeV.} 
\end{eqnarray}

Those numbers can be compared with Eqs.~\ref{eq:LimHB}, ~\ref{eq:LimHWB} and  \ref{eq:LimHW}, corresponding to the limits from the contribution to dimension-six operators. As one would expect, for similar $c_X$ values, the dimension-six are stronger than dimension-eight.  Later, we will compare those dimension-six and dimension-eight contributions for the benchmark models we proposed.

After obtaining  the sensitivity from the LHC and deriving limits of the dimension-eight contributions, one can assess  the range of validity of this result. As any other effective field theory, the description of NTGCs in terms of dimension-six and -eight operators is an expansion on energy over scale, $\hat s/\Lambda^2$ and is only valid under certain assumptions. First, note that those limits in the scale of new physics are in the TeV range for values of $c$ of order 1, that we have made use of a high-mass bin to obtain them, e.g. $\hat s > (1.3$ TeV$)^2$ in the CMS analysis, and most events will be located right at the border $\sqrt{\hat s}\sim 1.3$ TeV. If we then impose that the EFT is a valid assumption in the range $s/\Lambda^2 < 1$, then the limits on Eq.~\ref{eq:limitscurrent} can approximately be translated into how small the couplings $c$ could be before invalidating the expansion. Indeed,
to use that high-mass bin, the coefficients $c_{XY}$ have to be larger than $\sim 5$.

Moreover, in specific models those coefficients are related to each other, and
to the masses and couplings in the new sector. We will discuss this
translation to models in the next section.

\subsection{Model interpretation}

As we have seen in the previous section, searches for two neutral gauge
bosons in the final state can be used to place limits on NTGCs. Those
anomalous couplings are generated in the SMEFT as a combination of the
dimension-eight coefficients $c_{DAB}$ and the scale $\Lambda$, and we
showed that the Run2 limits on individual $c_{DAB}$, see
figure \ref{fig:ZZcurrent}, are of order one for $\Lambda=$ 1 TeV. In
section~\ref{sect:models} we have connected these operators in terms of
UV models and propose several benchmarks to test against experimental
results. Indeed, in table~\ref{tab:CXY_2VLF} we present a list of
models with two vector-like fermions and their contribution to
$c_{DAB}$. In the upper part of the table, we place the benchmark
$DS1$, a UV completion with a heavy lepton doublet and singlet. In the
lowest row, we place $QQ5$, a model which contains $SU(2)$ quintuplets
and quadruplets with exotic hypercharges and no color. Those two
models have the smallest and largest values of $c_{DAB}$ and provide a
good sample of the range of coefficients one should expect from the 2
VLF sets of models.

As can also be seen in the table, benchmarks do not produce individual
SMEFT coefficients, but a combination with specific sign relations,
and mass and coupling $Y$ dependence.  Those relations can be exploited to
place limits on UV models and connect with their discovery
potential. In other words, for each benchmark we can project the three dimensional $\chi^2(c_{D B \tilde B}/\Lambda^4, \, c_{D W \tilde B}/\Lambda^4, \, c_{D W \tilde W}/\Lambda^4)$ into a 1D $\chi^2(Y^2/m^2)$. 

In figure \ref{fig:massY}, we show how the SMEFT
dimension-eight limits translate into mass vs coupling regions for
models DS1 and QQ5. Despite the inherent loop suppression of these
contributions, current Run2 LHC limits reach masses of a few hundreds
of GeVs for the QQ5 benchmark and sizeable coupling, whereas for the
DS1 benchmark the reach is clearly outside of the range of validity of
the EFT.  Specifically, the limits at 95\% C.L. that we obtain scale
with the Yukawa coupling as:
\begin{equation}\label{eq:426}
\textrm{Run2 LHC: } m \gtrsim 190 \, (60) \times \, Y^{1/2} 
       \textrm{ GeV, for the QQ5 (DS1) model.}  
\end{equation} 
 
 Those limits are quite weak and could be possibly be surpassed by a direct search for pair-production of the new fermions through electroweak couplings. However, these direct searches depend on the fermion representation and the coupling of the new states to a SM fermion and gauge boson, a coupling which does not enter into the NTGC analysis and could be rather small. Depending on its value, the new fermions could be collider-stable, produce displaced vertices with leptons or prompt decays. The reinterpretation of various signatures, e.g. slepton searches, in terms of the benchmarks we have proposed goes beyond the scope of this paper.
 
 Finally, note that the limit from NTGCs in~\ref{eq:426} is competitive  with the limits we obtained from the re-interpretation of dimension-six global fits, Eq.~\ref{eq:LimHB2}. In the next section, we discuss how the current LHC limit will improve with luminosity.

\subsection{Prospects for High-Luminosity LHC}

\begin{figure}[t!]
\begin{center}
\includegraphics[width=4in]{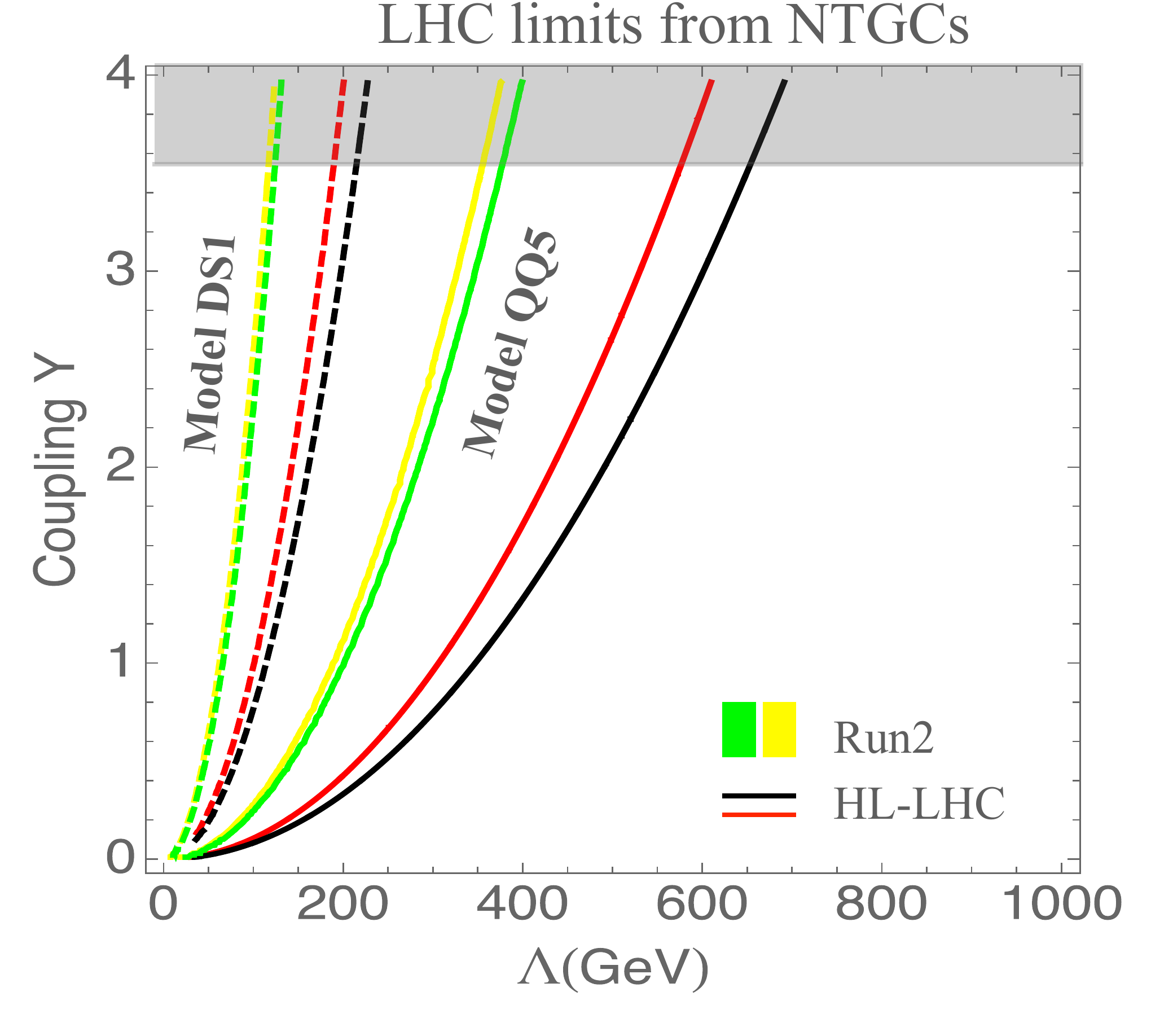}
\caption{Dependence of the NTGC limits on the mass of the new states
  $\Lambda$ versus coupling $Y$, for the illustrative model choices DS1
  (dashed) and QQ5 (solid). Green and yellow lines indicate the 1- and
  2-$\sigma$ current Run2 limits. The red (black) lines correspond to
  projections for discovery and exclusion at the HL-LHC,
  $\alpha=5$(2). The gray shaded region corresponds to the
  non-perturbative coupling region $Y>\sqrt{4 \pi}$. }
\label{fig:massY}
\end{center}
\end{figure}

In section~\ref{sec:NTGCpheno}, we examined two Run2 analyses (ATLAS and
CMS) and focused on the extremely clean $ZZ$ to 4L final state. In
both cases, the experiments provided differential distributions of the
4L invariant mass, with an estimate of the background in those
bins. Whenever possible, measurements with their errors were provided
by the experiments. As expected from the discussion around
figure~\ref{fig:distrib}, we found that the strongest bound was placed
through the bin with the highest mass. The dependence of those regions
on the dimension-eight coefficients vary as we move along the
distribution. In the Run2 ATLAS analysis~\cite{ATLAS:2021kog}, the
highest bin is at $m_{4\ell}>$ 900 GeV and the cross-section
dependence with, for example, $c_{D W \tilde W}$ is $\sigma\simeq 9.3 \,
(c_{D W \tilde W}/\Lambda^4)^2$ fb with $\Lambda$ in TeV. On the
other hand, CMS~\cite{CMS:2020gtj} provides information in a bin where
no events have been measured, but an estimate of the background
uncertainty was provided. For CMS, the bin was defined as $m_{4\ell}>$
1300 GeV, a region where the dependence on the dimension-eight
coefficient is still sizeable, $\sigma\simeq 4.8 \, (c_{D W \tilde
  W}/\Lambda^4)^2$ fb, with $\Lambda$ in TeV.

In this section we move from current limits to discuss projections for
the high-luminosity LHC (HL-LHC) run. As the CMS analysis shows, the
sign of new physics in NTGCs and its growth with parton
energy, could be explored in the extreme kinematic regions (typically,
the last bin), even if no events were measured.

To estimate the significance of a number $S$ of signal events in a
region that is not background ($B$) dominated, we are going to use a
statistical criteria based on the quantity~\cite{Bityukov:2000tt}
\begin{equation}
\alpha= 2 (\sqrt{S+B} - \sqrt{B}), \, 
\end{equation}
which reduces to the known $S/\sqrt{B}$~\cite{Feldman:1997qc} when
$S\ll B$, but is more robust under fluctuations and particularly
suited to experimental prospects in extreme kinematic
regions~\cite{Barducci:2015ffa}.

Based on this criteria, one can estimate the reach for exclusion
($\alpha=2$) or discovery ($\alpha=5$) with the full HL-LHC luminosity
of 3 ab$^{-1}$.  To illustrate the reach, we can take the CMS estimate
of 2.8 background events in the region $m_{4\ell}>$1300 GeV with Run2
full luminosity, our Monte-Carlo simulation of the signal and scale
up to HL-LHC. In figure~\ref{fig:massY}, we show the discovery and
exclusion potential of HL-LHC for the DS1 and QQ5 benchmarks. In
particular, for the QQ5 benchmark, the increase in luminosity brings
the reach closer to the TeV scale for ${\cal O}$(1)
couplings. Concretely, we estimate that the HL-LHC can exclude at 95\%
C.L. the following combination of masses and couplings,
\begin{equation}
\textrm{HL-LHC: } \Lambda \gtrsim 350  \,  (115) \times \, Y^{1/2} \textrm{ GeV, for the QQ5 (DS1) model.}  
\end{equation}
As discussed in section~\ref{sect:models}, predictions for models with 2
VLF will lie somewhere in between these two extremes. Also, note that
we have presented prospects using a single bin from the CMS analysis
and scaling it up with luminosity. In the future, one could improve
this analysis by including prospects for more channels (e.g. the
$Z\gamma$ or semileptonic $ZZ$), more bins in the invariant mass
(e.g. a few more bins from 1.3 TeV) and more observables in each
channel (e.g. sphericity or $\Delta R_{\ell^+\ell^-}$) provided the
correlations are taken into account.


\section{Conclusions\label{sec:cncl}}

NTGCs provide important tests for the gauge structure of the standard
model and are actively searched for by the main LHC collaborations.
In this paper, we have discussed standard model extensions with heavy
vector-like fermions as the simplest possibility to generate anomalous
neutral triple gauge boson couplings. We have generated a (largish)
list of different models and given their matching, both to SMEFT 
operators and to the experimentally measurable NTG couplings. 

In SMEFT, NTGCs are generated at the level of $d=8$ operators and we
have dedicated a section of this paper to discuss the relation between
the relevant operators and the vertices that experimentalists actually
can measure. We have shown that {\em four different} $d=8$ operators
are needed for a complete matching of models to the measurable form
factors. We have also discussed that the CPC $d=8$ operator chosen in
\cite{Degrande:2013kka} as a basis for NTGCs does not cover all
experimentally relevant vertices and thus is not sufficient for an
accurate experimental analysis.

Finally, we have estimated limits on the scale of the new physics
models from experimental data on the four-lepton $ZZ$ final state by
ATLAS~\cite{ATLAS:2017bcd} and CMS~\cite{CMS:2020gtj}. We interpreted
it in terms of NTGCs and translated these limits in terms of mass and
couplings of new VLFs. We found that the current limits are quite
weak, e.g. for the QQ5 benchmark and coupling $Y=\sqrt{4 \pi}$, the
mass limit is around 350 GeV.  Nonetheless, NTGC searches are
currently not background limited, and one can expect sizeable
improvements as more data is gathered. Indeed, our projections show
that for the same benchmark, one could reach scales close to the TeV in
the high-luminosity phase of the LHC.

\section*{Acknowledgements}

We would like to thank José Santiago for help with the tool
MatchMakerEFT~\cite{Carmona:2021xtq} and Javier Fuentes-Martín, Julie
Pagès and Anders Eller Thomsen for help with Matchete
\cite{Fuentes-Martin:2022jrf}.\\
F.E. is supported by the Generalitat Valenciana under the grants
GRISOLIAP/2020/145 and PROMETEO/2021/083. The research of V.S. is
supported by the Generalitat Valenciana PROMETEO/2021/083 and the
Ministerio de Ciencia e Innovacion
PID2020-113644GB-I00. M.H. acknowledges support by grants
PID2020-113775GB-I00 (AEI/10.13039/ 501100011033) and CIPROM/2021/054
(Generalitat Valenciana).  R.C. is supported by
MCIN/AEI/10.13039/501100011033 and the European Union
NextGenerationEU/PRTR under the grant JDC2022-048687-I and partially
funded by Grant AST22\_6.5 (Consejería de Universidad, Investigación e
Innovación and Gobierno de España and Unión Europea –
NextGenerationEU).

\bibliographystyle{JHEP}
\bibliography{dim8-ZZZ}

\end{document}